\documentclass[
  journal=largetwo,       
  manuscript=article-type, 
  year=2020, 
  volume=37,
]{cup-journal}

\usepackage[utf8]{inputenc}  
\usepackage{amsmath}         
\usepackage[nopatch]{microtype}  
\usepackage{booktabs}        
\usepackage{url}             
\usepackage{threeparttable}  
\usepackage{float}           
\usepackage{subcaption}      
\usepackage{orcidlink}       

\DeclareUnicodeCharacter{03C0}{$\pi$}

\newcommand{\apjl}{The Astrophysical Journal}
\newcommand{\aap}{Astronomy \& Astrophysics}

\newcommand{\apjs}{The Astrophysical Journal Supplement Series}

\newenvironment{acknowledgements}
  {\section*{Acknowledgements}} 
  {}

\title{VLBA astrometry of PSRs B0329+54 and B1133+16: Improved pulsar distances and comparison of global ionospheric models}

\author{Ashish Kumar\orcidlink{https://orcid.org/0009-0002-5290-037X}}
\affiliation{Department of Physics, Indian Institute of Technology, Kanpur-208016, India}
\email[Ashish Kumar]{kalyanaastro@gmail.com}

\author{Adam T. Deller\orcidlink{https://orcid.org/0000-0001-9434-3837}}
\affiliation{Centre for Astrophysics and Supercomputing (CAS), Swinburne University of Technology, John St, Hawthorn, VIC 3122, Australia}

\author{Pankaj Jain\orcidlink{https://orcid.org/0000-0001-8181-5639}}
\affiliation{Department of Space, Planetary $\&$ Astronomical Sciences $\&$ Engineering (SPASE), Indian Institute of Technology, Kanpur-208016, India}

\author{Javier Mold\'{o}n\orcidlink{https://orcid.org/0000-0002-8079-7608}}
\affiliation{Jodrell Bank Centre for Astrophysics, School of Physics and Astronomy, The University of Manchester, Manchester M13 9PL, UK}
\alsoaffiliation{Instituto de Astrof\'isica de Andaluc\'ia (IAA-CSIC), Glorieta de la Astronom\'ia s/n, E-18008 Granada, Spain}

\keywords{parallaxes - proper motions - pulsars:individual (PSR B0329+54, PSR B1133+16) - techniques:interferometric} 

\begin{document}

\begin{abstract}
Very long baseline interferometry (VLBI) astrometry is used to determine the three-dimensional position and proper motion of astronomical objects. A typical VLBI astrometric campaign generally includes around ten observations, making it challenging to characterise systematic uncertainties. Our study on two bright pulsars, B0329+54 and B1133+16, involves analysis of broadband Very Long Baseline Array (VLBA) data over $\sim30$ epochs (spanning approximately 3.5\,years). This extended dataset has significantly improved the precision of the astrometric estimates of these pulsars. Our broadband study suggests that, as expected, the primary contribution to systematic uncertainties in L-band VLBI astrometry originates from the ionosphere. We have also assessed the effectiveness of the modified TEC (total electron content) mapping function, which converts vertical TEC to slant TEC, in correcting ionospheric dispersive delays using global TEC maps. The astrometric parameters (parallax and proper motion) obtained from the multiple data sets, calibrated using the traditional and the modified TEC mapping functions, are consistent. However, the reduced chi-square values from least-squares fitting and precision of the fitted astrometric parameters show no significant improvement, and hence, the effectiveness of the new TEC mapping function on astrometry is unclear. For B0329+54, the refined parallax estimate is $0.611^{+0.013}_{-0.013}$\,mas, with best-fit proper motion of $\mu_{\alpha} = 16.960^{+0.011}_{-0.010}\, {\rm mas\, yr^{-1}}$ in R.A. and and $\mu_{\delta} = -10.382^{+0.022}_{-0.022}\, {\rm mas\, yr^{-1}}$ in Dec. These correspond to a distance of $1.64^{+0.03}_{-0.03}$\,kpc and a transverse velocity of $\sim 154\, {\rm km\, s^{-1}}$. For B1133+16, the new estimated parallax is $2.705^{+0.009}_{-0.009}$\,mas, with proper motions of $\mu_{\alpha} = -73.777^{+0.008}_{-0.008}\, {\rm mas\, yr^{-1}}$ and $\mu_{\delta} = 366.573^{+0.019}_{-0.019}\, {\rm mas\, yr^{-1}}$, implying a distance of $370^{+1}_{-1}$\,pc and a transverse velocity of $\sim 656\, {\rm km\, s^{-1}}$. The proper motions of B0329+54 and B1133+16 are consistent within $1\sigma$ of the most precise values reported in the literature to date while achieving more than a twofold improvement in precision. Similarly, the parallax measurements for both pulsars show a $\sim 73\%$ enhancement in precision, with differences of approximately $< 1\sigma$ compared to the most precise published values to date.
\end{abstract}

\section{Introduction}
Determining astronomical distances has significant challenges, with different distance scales relying on different methods. The parallax method is a direct approach for distance estimation, forming the foundation of the so-called \texttt{distance ladder}, but its utility is limited to a few kpc due to the small parallax contribution at large distances. Very long baseline interferometry (VLBI) and Gaia astrometry are capable of measuring positions with a precision of 0.1\,mas or better, which extends the range of accurate distance measurements up to 10\,kpc. To obtain a significant parallax contribution for a nearby isolated astronomical object, observations at a minimum of three epochs over approximately 1-2\,years are required. The source positions obtained from these observations allow us to fit five free parameters, position (R.A. and Dec.), proper motion in R.A. and Dec., and parallax. In the case of a binary system, more number of measurements are required to estimate orbital parameters. 

VLBI is among the most precise tools for astrometry, which provides comparable astrometric precision to that obtained by Gaia for optically bright sources. Ground-based VLBI offers sub-mas resolution, which enables the determination of the precise positions of radio sources\footnote{
The position uncertainty is given by $\frac{1}{2}\frac{\Delta \theta}{\text{S/N}}$, where $\Delta \theta$ and S/N represent the half-power width of the synthesised beam and the signal-to-noise ratio achieved on the target, respectively.
\label{foot:1}}. To facilitate a precise parallax measurement, observations are often clustered during a time at which the parallax signature is most readily detectable. For arrays with a greater East-West extent than North-South, such as the Very Long Baseline Array (VLBA), the higher angular resolution in R.A. makes it advantageous to observe near the peak parallax displacement in R.A.-- this is particularly enhanced for pulsars near the ecliptic plane, where the parallax displacement in R.A. significantly exceeds that in Dec. 

Relative astrometry offers more accurate distances, determining the target position with respect to a background reference source at multiple epochs. In relative astrometry, the possible differences in the atmosphere sampled by the target and calibrator dominantly affect the precision of the target position, and hence, the parallax estimates. Fortunately, numerous calibration techniques have been developed to reduce its effect. The phase referencing technique \citep{1995ASPC...82..327B} eliminates a majority of phase errors by determining the calibration solutions on a phase calibrator and then applying them to the target by interpolating over time. The primary phase calibrator is often a few degrees from the target, and any difference in atmospheric contributions seen between them remains uncompensated, which can be however reduced using In-Beam Calibration, that is by considering a secondary phase calibrator (typically a few arc minutes from the target) in the primary beam itself \citep{Chatterjee_2009, Deller_2019}. Nowadays, advanced calibration techniques using one- and two-dimensional interpolation \citep{Fomalont_2003, Rioja_2017, 10.1093/mnras/staa2531} have been developed to provide refined calibration solutions on the target. The only problem in employing these advanced calibration techniques in practice is a paucity of reliable calibrators in the vicinity of the target. Recently, \cite{ding2024pinpt} have introduced an advanced calibration method named \texttt{PINPT}, which integrates two-dimensional interpolation for eliminating spatial residual delays and corrects for frequency-dependent \texttt{core-shifts} \citep{bartel_core_shift, lobanov_core} of the phase calibrator, which is also a dominant systematic error contributor, by utilising observations across multiple frequencies. This approach has enabled them to determine the position of PSR J2222-0137 with exceptional precision. A comprehensive review of radio astrometry and some of the mentioned calibration techniques can be found in \cite{rioja2020precise} and \cite{2022arXiv221208881D}.

Pulsar distances can also be estimated using dispersion measure (DM) values and a model of Galactic electron distribution (e.g., \citealt{2002astro.ph..7156C, Yao_2017}). The derived distances can be significantly uncertain if the Galactic electron distribution model is inaccurate or any discrete density enhancements along the line of sight are not accounted for along with usually assumed smooth trends in the model (for an excellent early example of the latter, see \cite{1969MNRAS.146..423P}, which carefully took into account the excess DM contribution along the line of sight of a pulsar by considering the Str\"{o}mgren sphere of O/B type stars near the line of sight). \cite{1998MNRAS.300..577D} introduced an alternative technique to estimate pulsar distances, utilising scintillation and proper motion measurements. However, this method requires information about the distance to the dominant scatterer to translate the estimate of the fractional scatterer distance to the pulsar distance. More recently, \cite{10.1093/mnras/stac096} also presented a method for estimating pulsar distances based on scintillation measurements, but both techniques require the distance to the dominant scatterer.

Canonical pulsars, isolated (non-binaries) pulsars with a period grater than 20\,ms, have steep spectra with an average spectral index of -1.6 \citep{10.1093/mnras/273.2.411, 10.1093/mnras/stx2476}, and hence the highest signal-to-noise ratio (S/N; and best statistical astrometric accuracy) can generally be obtained by observing at lower frequencies since the improved S/N more than compensates for the lower instrumental angular resolution ($\propto\nu^{-1}$). However, ionospheric dispersive delay effects decrease with increasing frequency ($\propto\nu^{-2}$). These opposing position accuracy trends with frequency suggest a sweet spot for pulsar astrometric observations, which is typically around 1-2\,GHz and depends on the pulsar flux and observing telescope. However, for very bright pulsars or very sensitive arrays, higher frequency observations may be more effective. The optimum frequency for pulsar observation also depends on the sophistication of the ionospheric mitigation techniques, as well as our knowledge of how well they work. The limited number of observations in the earlier VLBI astrometric campaigns makes it difficult to directly constrain the ionospheric systematic effects since the small number of degrees of freedom in astrometric fits, combined with other partially degenerate effects, such as reference source structure evolution, make direct estimates uncertain \citep{ding2021orbital}. Accordingly, astrometric campaigns with a larger number of observations, especially spanning multiple frequencies, are valuable for studying systematic uncertainties.

In this paper, we report on a set of multi-frequency observations of two radio bright pulsars observed at more than 30 epochs spanning three years, offering a chance to better estimate the astrometric parameters and characterise the various contributions to systematic uncertainties. The paper is structured as follows: Section \ref{sec:2} provides an overview of the VLBA observations of the BD174 data set and correlation details, while Section \ref{sec:3} outlines the data reduction pipeline and imaging recipe employed in the study. Astrometric fitting techniques are elucidated in Section \ref{sec:4}. The impact of total electron content (TEC) mapping function and multi-frequency study of ionospheric effects on astrometry are discussed in Sections \ref{sec:5} and \ref{sec:6}, respectively. The conclusions of this study are summarised in Section \ref{sec:7}.

\section{Observations and correlation}\label{sec:2}
The pulsar astrometric analysis presented here uses two VLBA data sets, project codes: 1) BD174, which is being analysed and presented for the first time, and 2) BD152, presented earlier in \cite{Deller_2019}. The BD152 pulsar astrometric campaign observed 57 pulsars at 1.6\,GHz in 8/9 epochs (similar to a typical VLBI astrometric campaign) spanning more than two years (from January 2011 to December 2013). BD174 extends this with another more than 20 observations of two bright pulsars, B0329+54 and B1133+16, lasting less than a year. The BD174 observing parameters are presented in Table \ref{tab:bd174_obs}. These observations were designed to better understand the contributions of systematic uncertainties to astrometric parameters, and were conducted in ``filler'' time, with short observing duration and (sometimes) fewer antennas available.
\begin{table}[htbp]
\caption{Details of the BD174 observations.}
\begin{tabular}{ll}
\toprule
Observation date        \dotfill &  October 2013 to October 2014    \\
Number of epochs        \dotfill &  29 (B0329+54) and 24 (B1133+16) \\
Epoch duration          \dotfill &  $\sim$ 1\,hr   \\
Polarization            \dotfill &  LL and RR     \\
Frequency bands         \dotfill &  L (1–2\,GHz) and S (2–4\,GHz) \\
L-band subband centre freqs (GHz)   \dotfill &  1.38, 1.41, 1.63, and 1.66\\
Bandwidth (per band)    \dotfill &  32\,MHz  \\
Frequency resolution    \dotfill &  500\,kHz   \\
Data rate               \dotfill &  1024\,Mbps   \\
RMS noise (L-band)      \dotfill &  $\sim 85\, \mu{\rm Jy\, beam}^{-1}$   \\
\bottomrule
\end{tabular}
\label{tab:bd174_obs}
\end{table}

\begin{figure}
    \centering
    \includegraphics[width=\linewidth]{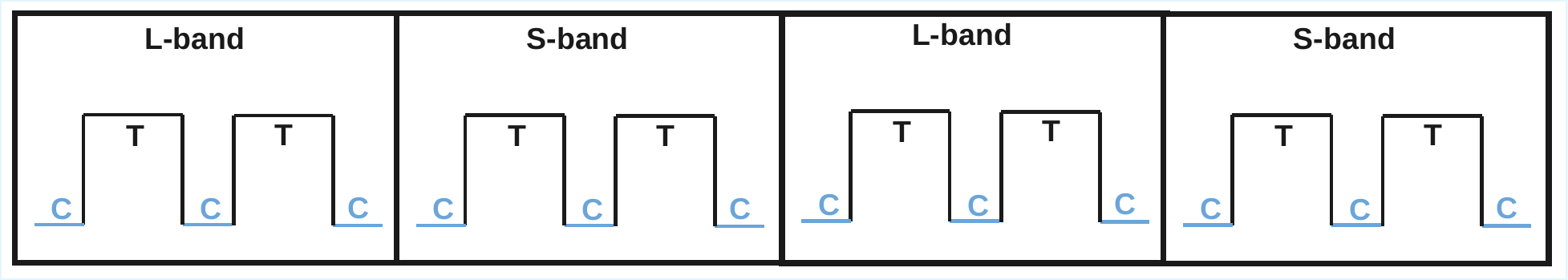}
    \caption{Illustration of the BD174 observing strategy. Each observation is divided into four blocks, two at L-band and two at S-band, as shown by rectangular boxes. Within each block, two target scans (labelled as ``T'') are observed in an interleaved manner, with each target scan bracketed by scans of the Phase Reference Calibrator (labelled as ``C'').}
    \label{fig:bd174_observings}
\end{figure}

The following description of observing strategy and parameters pertains only to the BD174 data. Each observation is broken into four blocks, where a block consists of two scans on the target field (each of five minutes) interleaved and bracketed by scans on the Phase Reference Calibrator (PRC; each one minute), which are followed by a Fringe Finder Calibrator (FFC; one minute) scan. The first block is at L band, followed by an S band block, a second L band block, and a second S band block (Figure \ref{fig:bd174_observings}). In total, $4 \times 5$\,minutes is thus obtained on the target field at each band. We have excluded the S-band data from the analysis since it has root-mean-square (RMS) noise almost ten times higher than the theoretical expectation due to the huge radio frequency interference (RFI) in this band (for details, see VLBA Scientific Memo 41\footnote{\url{https://library.nrao.edu/public/memos/vlba/sci/VLBAS_41.pdf}}). Consequently, our analysis focuses solely on the L-band data. The target field, with the pulsar positioned at the pointing centre, included three In-Beam Calibrators (IBCs). Details of observed pulsars with their period, DM, gated equivalent flux density at 1.4\,GHz, and calibrators with their separation from the pulsar, along with their flux densities at 1.4\,GHz and 1.6\,GHz for calibrators, are listed in Table \ref{table:pulsar_calibrators}.

The data were correlated using \texttt{DiFX} software correlator \citep{Deller_2007, Deller_2011}. Each observing epoch is processed through five correlation passes: three for IBCs and two for pulsar, one in ungated and the other in gated pulsar mode\footnote{The pulsar gating, by discarding off-pulse data, enhances the S/N  by a factor of $f^{-1/2}$, here, $f = \frac{T_{\text{on}}}{T_{\text{on}} + T_{\text{off}}}$ represents the duty cycle of the pulsar, where $T_{\text{on}}$ and $T_{\text{off}}$ denote the on-pulse and off-pulse time intervals, respectively.\label{foot:4}}.
For pulsar gating, the \texttt{DiFX} software correlator utilises a ``polyco'' file, which contains pulse longitude information as a function of frequency and observation time. For this astrometric campaign, the ``polyco'' files were generated utilising timing data obtained using the Lovell Telescope. The first IBC pass also includes processing of PRC and FFC scans. The correlation positions of the PRCs, listed in Table \ref{table:prc}, are taken from the ``.sum'' files\footnote{\url{http://www.vlba.nrao.edu/astro/VOBS/astronomy/}}. The table also includes the updated positions of these sources, taken from the Radio Fundamental Catalogue (RFC; \citealt{petrov2024_radio}). Our PRCs and IBCs do not have obvious calibrator structural variability on the scale of a significant fraction of the synthesised beam or larger.
\begin{table*}[htbp]
    \centering
    \resizebox{\textwidth}{!}{%
    \begin{threeparttable}
    \begin{tabular}{ccccccccccccc}
    \toprule
    \addlinespace
    Pulsar & Period\tnote{1} & DM\tnote{1} & F$_{1.4}$\tnote{2} & FFC & PRC & Sep\tnote{3} & \multicolumn{2}{c}{F$^{\text{PRC};}_{\nu}$\tnote{5}} & Reference frame & Sep\tnote{4} & \multicolumn{2}{c}{F$^{\text{IBC;}}_{\nu}$\tnote{5}} \\
    \cmidrule(lr){8-9} \cmidrule(lr){12-13}
    & & & & & & & 1.4\,GHz & 1.6\,GHz & & & 1.4\,GHz & 1.6\,GHz \\
    & (s) & (pc\,cm$^{-3}$) & (mJy) & & & (deg) & (mJy) & (mJy) & & (arcmin) & (mJy) & (mJy) \\
    \midrule
    \addlinespace
    B0329+54 & 0.71 & 26.76 & 1244.9 & J0319+4130 & J0346+5400 & 2.06 & 187.0  & 182.7 & J033317+544011 & 6.07 & 9.1  & 8.3 \\
    B1133+16 & 1.19 & 4.84  & 181.9  & J0927+3902 & J1142+1547 & 1.46 & 75.9  & 93.4  & J113609+155228 & 1.91 & 12.9 & 16.2 \\
    \bottomrule
    \end{tabular}
    \begin{tablenotes}
    \item[1] \url{https://www.atnf.csiro.au/research/pulsar/psrcat/}; \citep{Manchester_2005}.
    \item[2] \cite{Deller_2019}.
    \item[3] Angular separation between the target and the PRC.
    \item[4] Angular separation between the target and the IBC.
    \item[5] Flux densities estimated using JMFIT on global model images of the PRC and IBC, obtained by combining all epochs.
    \end{tablenotes}
    \caption{Parameters of target pulsars, as well as their associated bandpass and phase reference calibrators.}
    \label{table:pulsar_calibrators}
    \end{threeparttable}
    }
\end{table*}

\begin{table*}[htbp]
    \centering
    \begin{threeparttable}
    \begin{tabular}{p{1cm} p{1.5cm} p{2cm} p{1cm} p{2cm} p{1cm} p{2cm} p{0.8cm} p{2cm} p{0.8cm}}
    \toprule
    Pulsar & PRC & $\alpha$\tnote{1} & $\sigma_\alpha$\tnote{1} & $\delta$\tnote{1} & $\sigma_\delta$\tnote{1} & $\alpha$\tnote{2} & $\sigma_\alpha$\tnote{2} & $\delta$\tnote{2} & $\sigma_\delta$\tnote{2} \\
    \addlinespace
     & & (h:m:s) & (mas) & (d:m:s) & (mas) & (h:m:s) & (mas) & (d:m:s) & (mas) \\
    \midrule
    B0329+54 & J0346+5400 & 03:46:34.50413 & 0.41 & 54:00:59.10868 & 0.33 & 03:46:34.50414 & 0.14 & 54:00:59.10919 & 0.10\\
    B1133+16 & J1142+1547 & 11:42:07.73597 & 1.72 & 15:47:54.17676 & 3.71 & 11:42:07.73592 & 0.13 & 15:47:54.17902 & 0.20\\
    \bottomrule
    \end{tabular}
    \begin{tablenotes}
    \item[1] PRCs coordinates are taken from ``.sum'' files.
    \item [2] PRCs coordinates are taken from RFC \citep{petrov2024_radio}.
    \end{tablenotes}
    \caption{Correlation and RFC positions of the PRCs.}
    \label{table:prc}
    \end{threeparttable}
\end{table*}

\section{Data reduction and imaging}\label{sec:3}
The visibility data, resulting from \texttt{DiFX} processing, have been analysed primarily using \texttt{AIPS} (version 31DEC23; \citealt{2003ASSL..285..109G}), utilising \texttt{ParselTongue} \citep{2006ASPC..351..497K}, a Python interface, that facilitates \texttt{AIPS} functionality. The calibration pipeline, VLBI Data Calibration and Imaging pipeline in Python (VDCIpy)\footnote{
\url{https://github.com/kalyanastro/VDCIpy}\label{foot:6}}, 
is used in the present work (similar to \citealt{Deller_2019}). A brief summary of the calibration pipeline, including \texttt{AIPS} tasks, is as follows. It first performs a priori calibration, including ionospheric dispersive delay and Faraday rotation corrections using TEC maps (TECOR) and adjustments for the Earth orientation parameters using the latest Earth Rotation Parameter (ERP) file (CLCOR). The visibility amplitude corrections are applied through ACCOR and APCAL, which do the quantisation correction and convert the cross-correlation coefficients to Jy-unit, respectively. The primary beam corrections are performed using a \texttt{ParselTongue} script, which compensates for primary beam attenuation and beam squint \citep{middelberg2011_widefield}. The single-band delay, phase, and bandpass calibration solutions are derived using scans of the calibrators, listed in Table \ref{table:pulsar_calibrators}. Bandpass and time-independent single-band delay solutions are determined on the FFC scan using the BPASS and FRING tasks, respectively, and applied to all sources. The fringe-fitting solutions are computed on each band and on each scan of the PRC. Phase and amplitude self-calibration solutions are also obtained from PRC scans using the CALIB task. The calibration solutions obtained on the PRC are applied to both the PRC and target field sources. A calibrator model is used while performing the BPASS/FRING/CALIB task, which is generated by concatenating all epochs at the respective frequency bands.

To minimise the systematic errors introduced by the difference in the atmosphere sampled by the target and the PRC, we employ the inverse in-beam phase referencing technique \citep{10.1093/pasj/64.6.142, Deller_2019} since the target pulsars are adequately bright and can serve as the secondary phase calibrator. This technique determines the residual delays on the pulsar by incorporating the phase self-calibration (CALIB task, with a pulsar model constructed after primary calibration) and applies them to the IBC. To keep the residual delays at a minimum and avoid phase wraps, we have shifted the pulsar model position at each epoch based on the preliminary model of the pulsar motion. The preliminary model accounts for the pulsar position shift due to its proper motion, which can be taken from pulsar timing measurements or previous astrometric studies. The model position of the pulsar at $i^{th}$ epoch is $\text{P}_i^j\ =\ \text{P}_0^j\ +\ \mu_j t$, where $\text{P}_0$ is the position at the model epoch, $t$ is the time since the model epoch, and $j$ represents either R.A. or Dec.

For each pulsar, the first epoch of the BD174 campaign is excluded from the analysis since it does not have the FFC scan. To study the frequency-dependent effects, we have analysed the 1.4\,GHz and 1.6\,GHz data sets independently. For the 1.6\,GHz data, the BD152 calibrator models are used in calibration, as both are observed at nearly the same central frequency (1.651\,GHz versus 1.646\,GHz). The average beam sizes during the BD174 observations of B0329+54 and B1133+16 are mentioned in Table \ref{table:beam_size}. At some epochs, the beam is even broader, especially when a few antennas are missing during the observation or flagged in the calibration process. This can be due to the RFI or poor bandpass solutions at that antenna. 
\begin{table}[htbp]
    \centering
    \begin{threeparttable}
    \begin{tabular}{p{2.5cm} p{2.5cm} p{2.5cm}}
    \toprule
    Pulsar & $\theta_{1.4}$ & $\theta_{1.6}$ \\
     & (mas) & (mas)\\
    \midrule
    B0329+54 & 17.4 $\times$ 6.3 & 14.4 $\times$ 5.3 \\
    B1133+16 & 18.3 $\times$ 6.6 & 15.1 $\times$ 5.8 \\
    \bottomrule
    \end{tabular}
    \caption{Average synthesised beam size ($\theta$) during the BD174 observations at 1.4\,GHz and 1.6\,GHz.}
    \label{table:beam_size}
    \end{threeparttable}
\end{table}

The frequency-averaged calibrated visibility data for all sources are stored. Particularly, the in-beam calibrator u-v data is divided by its model to convert it to a pseudo-point source to improve the S/N. After that, the Stokes \textit{I} visibility data for all sources are Fourier transformed using \texttt{DIFMAP} (version 2.5p; \citealt{1997ASPC..125...77S}). The source images have been created using natural weighting, and are of a size of $1024 \times 1024$\,pixels with each pixel size of 0.75\,mas and 0.85\,mas for 1.6\,GHz and 1.4\,GHz data sets, respectively. The clean image of the IBC at each epoch is loaded into \texttt{AIPS} to fit an elliptical Gaussian centred at the peak using the JMFIT task. The position of the source can be determined by fitting a model directly to the visibility data. However, the reliability of the position uncertainties in this approach depends heavily on the scaling of the visibility weights, which may not fully account for calibration errors. On the other hand, image-plane fitting is insensitive to any constant scale factor errors in the weighting. For this reason (and to provide consistency with previous analyses of this VLBI data on these pulsars), we opted for image-domain fitting over visibility-domain fitting.

In the case of inverse phase referencing, we get a position-time series of the IBC, which is converted to the pulsar position-time series by subtracting the IBC position offset (P$^i_\text{IBC}$ - P$^0_\text{IBC}$; IBC position at i$^{th}$ epoch minus the mean position of the IBC) from the pulsar model position, and stored in \texttt{pmpar.in.preliminary} file\footnote{\url{https://github.com/kalyanastro/researchMaterials}\label{foot:9}}. 
The mean IBC position is estimated from its position-time series, which is obtained from the calibrated IBC data set, without applying pulsar self-calibration solutions. The \texttt{pmpar} file includes only the thermal uncertainties to the IBC positions. The pulsar model position uncertainty should also be added in quadrature, but since our pulsars are bright, their contribution to the overall positional uncertainties is negligible. Just for completeness, one can also fit astrometric parameters directly to the IBC position-time series, and can then force the in-beam source to have zero proper motion and parallax, which corresponds to subtracting its proper motion and parallax from the a priori pulsar model.

\section{Astrometric fitting}\label{sec:4}
\begin{figure*}[htbp]
    \centering
    \includegraphics[width=0.49\linewidth]{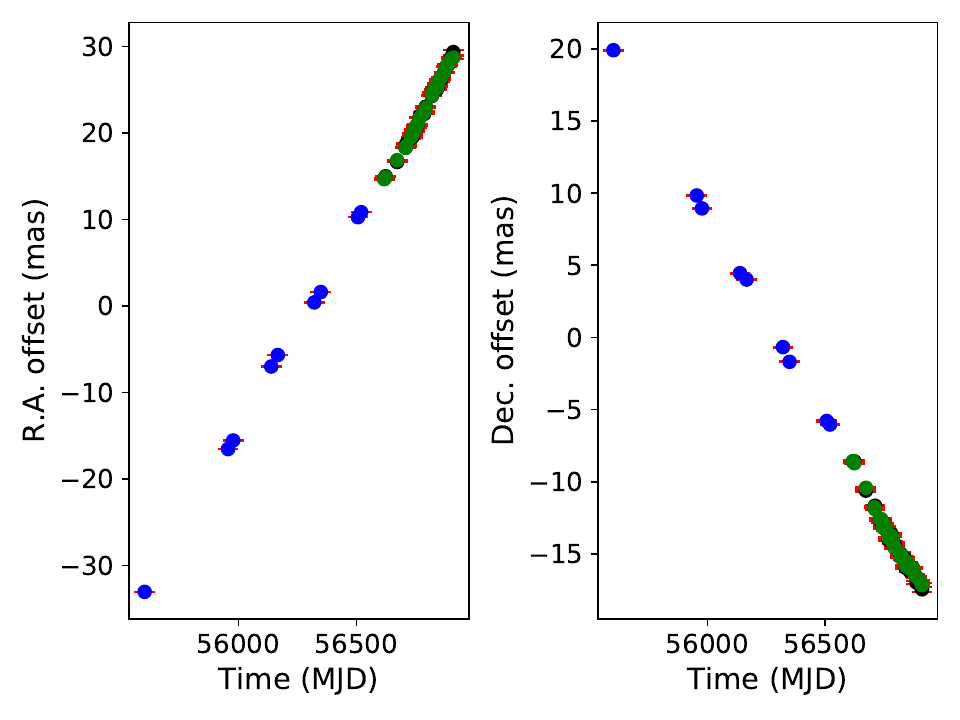}
    \includegraphics[width=0.49\linewidth]{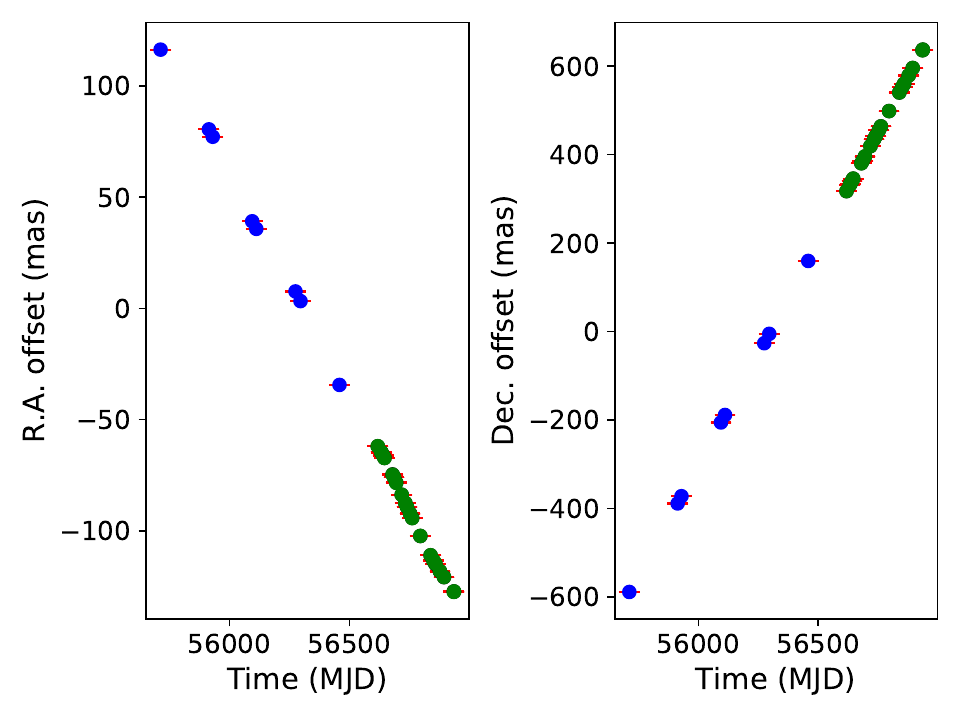}
    \caption{Position evolution of B0329+54 (left two plots) and B1133+16 (right two plots), including error bars solely based on thermal noise. Data from BD152 are shown in blue, while BD174 data are represented by black (1.4\,GHz) and green (1.6\,GHz). The plotted positions include the contributions from both the proper motion and the parallax. The best-fit reference epoch (MJD: 56300) position is subtracted from the pulsar position-time series.}
    \label{fig:psr_position_evolution}
\end{figure*}

For each pulsar, three position-time series are obtained: two from the BD174 data set at 1.4\,GHz and 1.6\,GHz and one from the BD152 data set (Figure \ref{fig:psr_position_evolution}). These position-time series are phase-referenced to the nearest IBC (Table \ref{table:pulsar_calibrators}) to minimise ionospheric effects. However, the reference sources—IBCs, typically Active Galactic Nuclei (AGNs)—exhibit a frequency-dependent position shift known as the \texttt{core-shift} \citep{bartel_core_shift, lobanov_core}. 
If a constant source model is assumed, this shift introduces a systematic offset between the 1.4\,GHz and 1.6\,GHz position-time series and propagates to pulsar positions. More generally, any difference between the actual and the modelled angular brightness distribution of the calibrator source at any frequency introduces a position shift in the calibration terms at that frequency that propagates to the pulsar position.

To correct for this misalignment, we adjust the 1.4\,GHz position-time series by subtracting the reference epoch positions offset (P$_{\text{ref}}^{1.4}$ - P$_{\text{ref}}^{1.6}$) from it. This approach is justified because 1.6\,GHz astrometry generally provides better position accuracy due to reduced ionospheric distortions and improved resolution. The reference epoch is chosen in the middle of the astrometric campaign to minimise the propagation of proper motion uncertainties to the reference epoch position. The reference epoch positions of B0329+54 and B1136+16 at 1.4\,GHz and 1.6\,GHz differ by approximately 2.2\,mas and 2.3\,mas, respectively. This position difference arises due to the independent phase referencing at each frequency. Once aligned, the corrected BD174 position-time series are merged with BD152 data to determine the final astrometric parameters. Since the 1.6\,GHz and BD152 data sets were observed at nearly the same frequency and calibrated using identical calibrator models, as expected, no significant systematic frequency-dependent offset was detected. Nevertheless, we applied the same merging procedure as used for 1.4\,GHz and 1.6\,GHz. The time offset between the BD174 and BD152 observations implies that any systematic frequency-dependent offset between the calibrator model used in BD152 and that in the 1.6\,GHz observations would mostly be absorbed into the proper motion fit. However, since the 1.6\,GHz and BD152 data sets were observed at nearly the same frequency, using an identical model is unlikely to introduce significant unmodeled core-shift errors.

During the astrometric parameters fitting, we have excluded a few epochs, particularly one epoch ($ae$) for B0329+54 and three epochs ($bk$, $bl$, and $bm$) for B1133+16, because they exhibit significant deviations from the model (described by Eq. \ref{eq:astrometric_motion}). Although we could not identify the exact reason, but as a part of our diagnostic process, we have calculated the summed chi-square values ($\displaystyle\sum_{i=1}^{3} \dfrac{(\text{P}_i - \text{P}_{\text{model}})^2}{\sigma_i^2}$, $\text{P}_{\text{model}}$ is expressed by Equation (\ref{eq:astrometric_motion})) for each epoch, considering all three IBCs. The summed chi-square values are around 40, 60, 250, and 130 for the mentioned epochs in that order, while the average chi-square value for the rest of the epochs is around five. The very large number of epochs in the BD174 observing campaign makes it feasible to exclude a few epochs, while it would generally not be feasible for a campaign with a typical number of epochs, $\sim10$. The final position-time series from BD174 and BD152 data sets (phase-referenced to the closest IBC), excluding the mentioned epochs, are shown in Figure \ref{fig:psr_position_evolution}.

We have employed three statistical inferences to estimate five astrometric parameters (position, $\alpha_{\text{J2000}}$, $\delta_{\text{J2000}}$, proper motion in R.A. and Dec., $\mu_{\alpha}$, $\mu_{\delta}$, and parallax, $\varpi$) from the concatenated position-time series:

1) \textbf{Least-square fitting:}\label{sec:4.1}
The \texttt{PMPAR}\footnote{\url{https://github.com/walterfb/pmpar}} (version 1.8) software is used to estimate astrometric parameters that are best fit to the position-time series of our pulsars based on a model of the apparent path of a star, which can be modelled as,
\begin{equation}
\left\{
\begin{aligned}
\alpha(t) &= \alpha_{\text{J2000}} + \mu_\alpha t + \varpi f_\alpha(t; \alpha_{\text{J2000}}, \delta_{\text{J2000}}) \\
\delta(t) &= \delta_{\text{J2000}} + \mu_\delta t + \varpi f_\delta(t; \alpha_{\text{J2000}}, \delta_{\text{J2000}}),
\end{aligned}
\right.
\label{eq:astrometric_motion}
\end{equation}
where $t$ is the time since the reference epoch, ($\alpha_{\text{J2000}}$, $\delta_{\text{J2000}}$) are the position of the pulsar at that epoch, and $f_\alpha$, $f_\delta$ represent the components of the apparent elliptical path associated with parallax in units of 1\,arcsec (which is the parallax value for a source at one pc distance) in R.A. and Dec., respectively,
\[
\left\{
\begin{aligned}
f_\alpha &= \dfrac{1}{15}\sec\delta (X\sin\alpha - Y\cos\alpha) \\
f_\delta &= X \cos\alpha \sin\delta - Y \sin\alpha \sin\delta - Z \cos\delta,
\end{aligned}
\right.
\]
where $X$, $Y$, and $Z$ are the epoch-dependent coordinates of the Earth's centre in the equatorial coordinate system, with its origin at the Solar system barycentre \citep{1985spas.book.....G}.

The parameter estimation using the least-squares method is always strongly influenced by the uncertainties in the input data. The reduced chi-square for astrometric parameters fitting generally exceeds unity (for example, the reduced chi-square is eleven for the data set on B0329+54), indicating a significant underestimation of position uncertainties. In the JMFIT task, the estimation of position uncertainties is based solely on thermal noise in the image. However, the total error budget should include systematic errors like atmospheric contributions, calibrator structural evolution, if any, etc. The atmosphere relevant to us comprises the troposphere and ionosphere, which introduce the nondispersive ($\propto\nu$) and dispersive ($\propto\nu^{-1}$) phase errors, respectively. The dominant systematic error contribution in L-band astrometry comes from the ionosphere \citep{Chatterjee_2004, Brisken_2000}, which mainly depends on the separation between the phase calibrator and the target \citep{Chatterjee_2004, Deller_2019}, emphasising the importance of choosing a suitable \texttt{reference frame} (IBC) close to the target. 

\cite{Deller_2019} have proposed the following empirical relation to compute the systematic error based on calibrator-target separation ($s$; in arcminutes), antenna elevation ($el$) during the observation, and S/N on the calibrator ($S$),
\begin{equation}
     \sigma_{\text{sys}} = A \times \dfrac{s \times \sum_{a=1}^{N_{\rm ant}} \sum_{o=1}^{M} \csc(el_{a, o})}{M \times N_{\rm ant}} + \dfrac{B}{S},
    \label{eq:sys_err_formula}
\end{equation}
here $\sigma_{\text{sys}}$ is the systematic error in units of the width of the synthesised beam; $el_{a,o}$ is the elevation of antenna $a$ in scan $o$ on the target pulsar; $N_\text{ant}$ and $M$ are the number of antennas and number of target scans, respectively and coefficients $A$ and $B$ are 0.001 and 0.6 respectively. These coefficients are estimated using \texttt{PSRPI} observations \citep{Deller_2019}. Therefore, the empirical relation provides the best estimate of the systematic errors at a mean frequency of 1.6\,GHz and the typical ionospheric condition similar to that time. We estimate the systematic error using Equation (\ref{eq:sys_err_formula}), add it to the random error in quadrature, and store it in \texttt{pmpar.in} file$^{\ref{foot:9}}$. The position uncertainty estimates then become more realistic, and the reduced chi-square, as expected, approaches unity. We find that in the case of B0329+54, the reduced chi-square turns out to be unity, while for B1133+16, the reduced chi-square is improved from twelve to three. The least-squares fitting estimates for our pulsars are mentioned in Table \ref{table:ast_results_56000}.

2) \textbf{Bootstrap fitting:}\label{sec:4.2}
As noted above, the least-squares fitting is not the optimal technique to estimate astrometric parameters, particularly their uncertainties, when uncertainties in input data are not properly accounted. In astrometry, systematic error contributions to the uncertainties in the measured positions are well appreciated but difficult to quantify. Hence, we appeal to the bootstrap fitting method \citep{doi:10.1126/science.253.5018.390}, which is less sensitive to input uncertainties and thus provides a more reliable estimate of the fitting parameters and their uncertainties. 

\begin{figure*}
  \centering
  \includegraphics[width=0.49\linewidth]{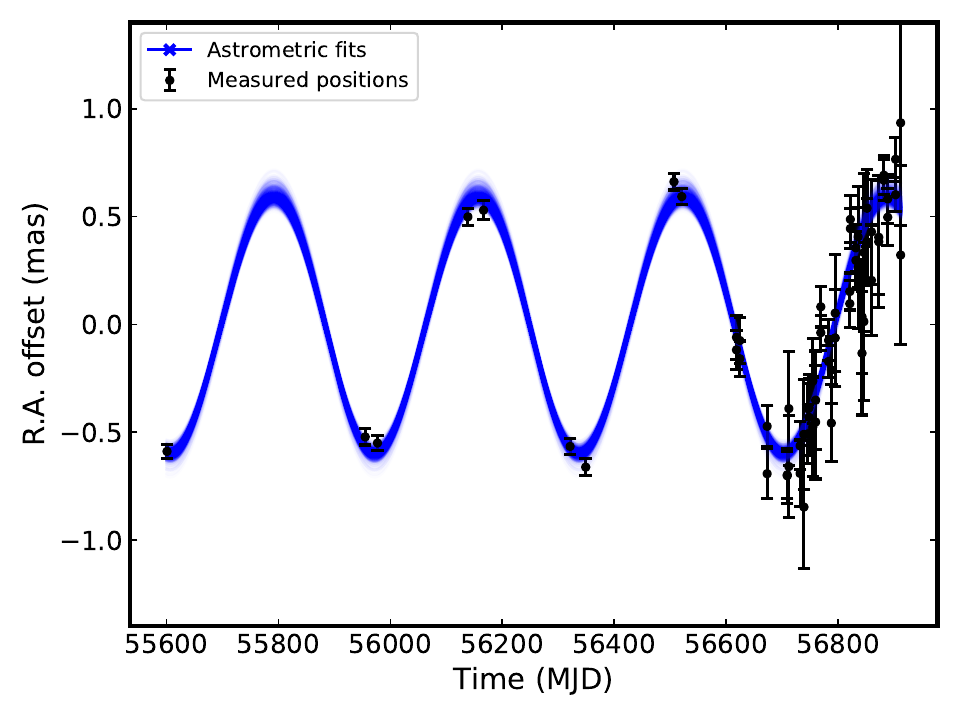}
  \includegraphics[width=0.49\linewidth]{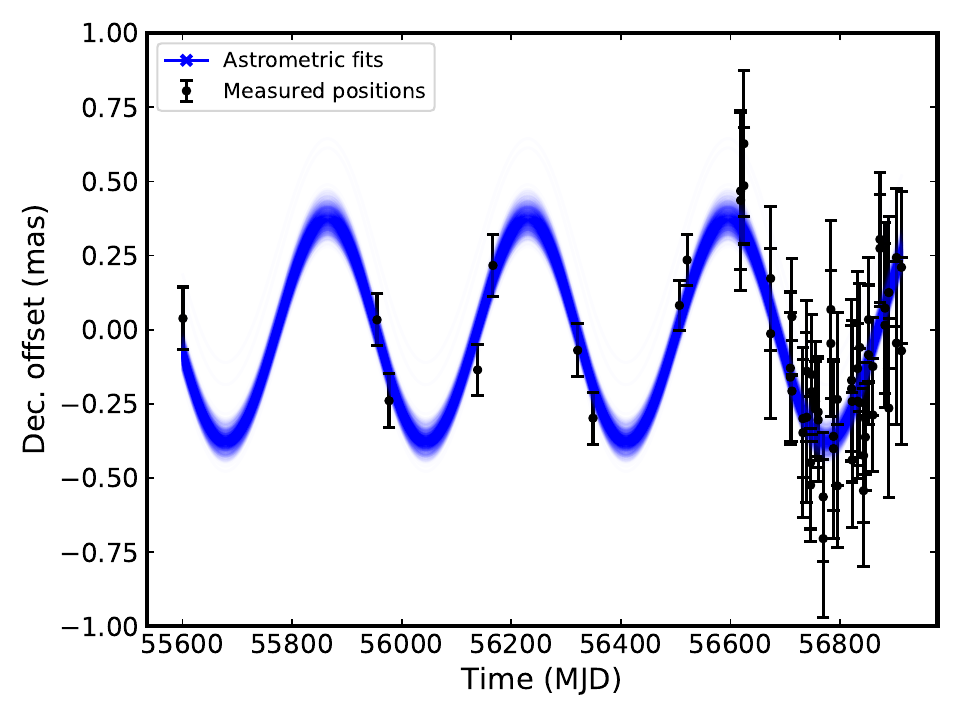}
  \includegraphics[width=0.49\linewidth]{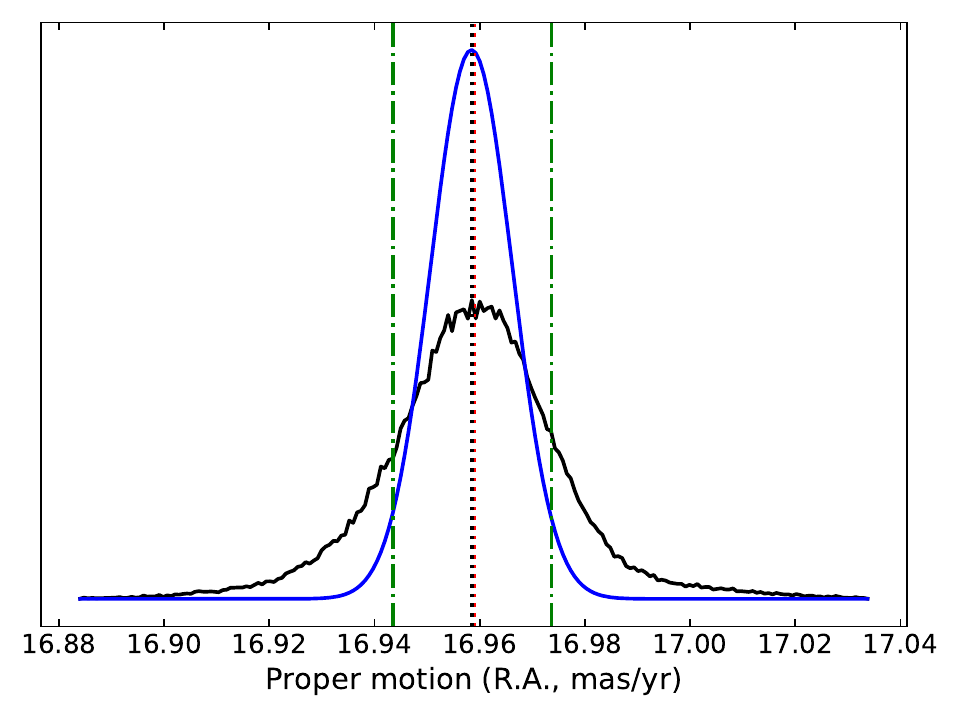}
  \includegraphics[width=0.49\linewidth]{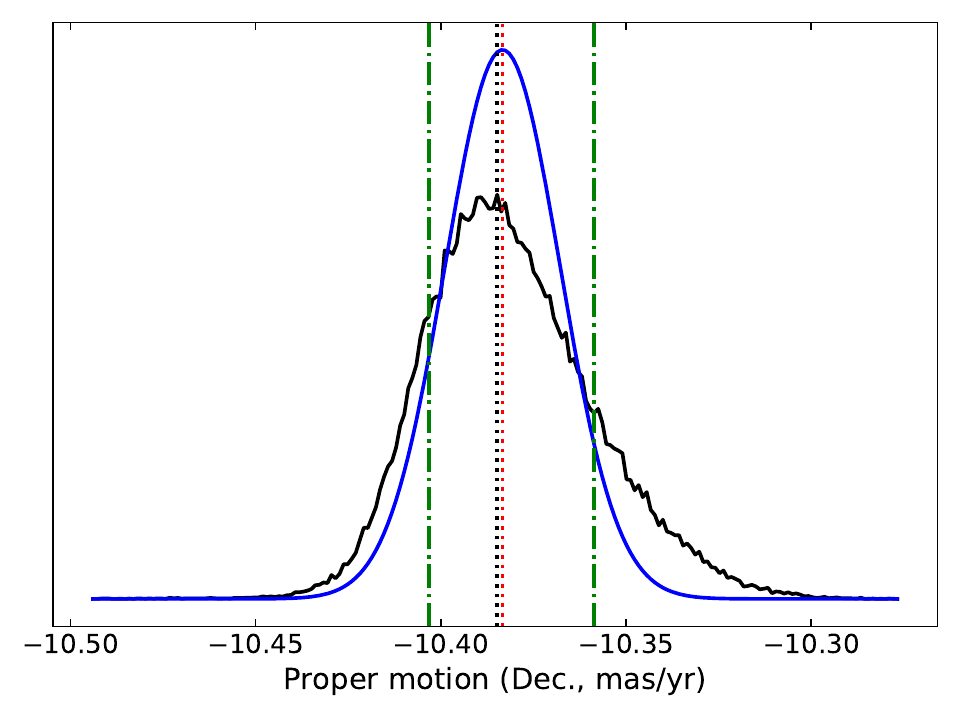}
  \includegraphics[width=0.49\linewidth]{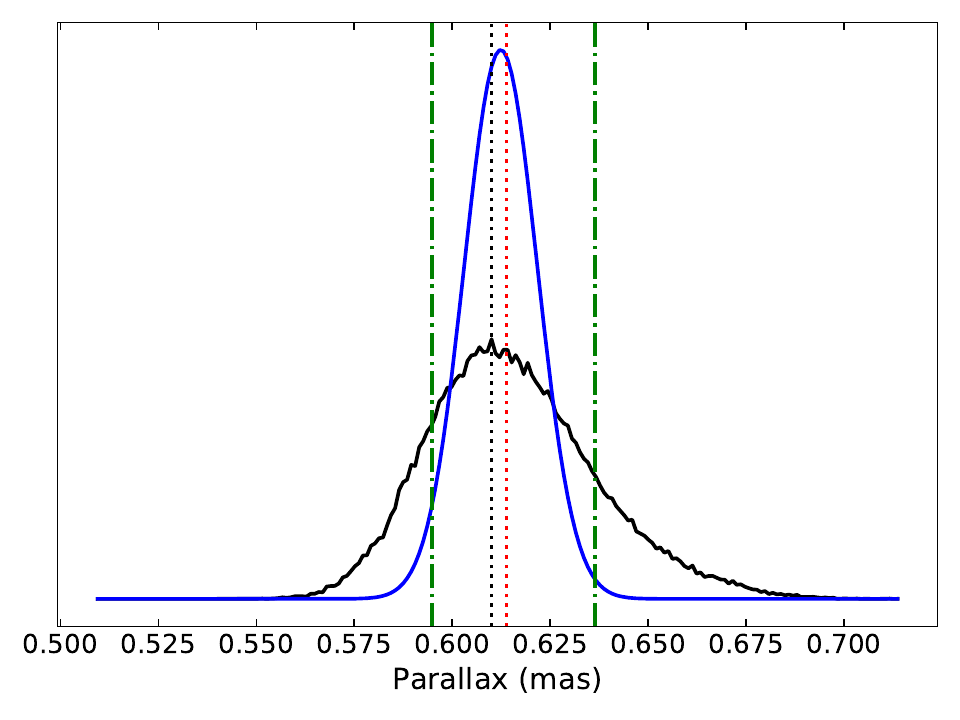}
  \caption{Illustration of bootstrap fitting estimates for B0329+54. Top: pulsar position offsets (R.A., left and Dec., right) from the reference epoch (MJD: 56300) position with error bars, plotted as a function of epochs after removing the best-fit proper motion. The first nine data points (from left) are from BD152 observations, while the remaining data points are from BD174 observations. Each BD174 epoch includes two data points, representing the 1.4\,GHz and 1.6\,GHz datasets. Each light blue line represents the parallax signature fitted to a respective bootstrap sample, and the band of these lines indicates the spread in the fitted parallax signature. The increased spread in the parallax signature curves for the Dec. offset, compared to the R.A. offset, can be attributed to a combination of a) sampling epochs that sample R.A. offsets closer to the parallax signature extremes in earlier epochs (i.e., BD152), and b) relatively larger error bars in Dec. offset estimates. Middle and bottom: probability density functions for the parallax and the proper motion in R.A. and Dec. are shown for both the bootstrap in black colour and least-squares results (Gaussian function with the parameters estimated using \texttt{PMPAR} after adding the systematic error contribution to obtain the reduced chi-square to be one, plotted in blue colour). The dotted vertical lines indicate the most probable value (black) and the median value (red), with green dashed-dotted lines marking the 68\% confidence interval ($\pm 1 \sigma$) estimated from the bootstrap fit.}
  \label{fig:boot_results_J0332}
\end{figure*}

\begin{figure*}
  \centering
  \includegraphics[width=0.49\linewidth]{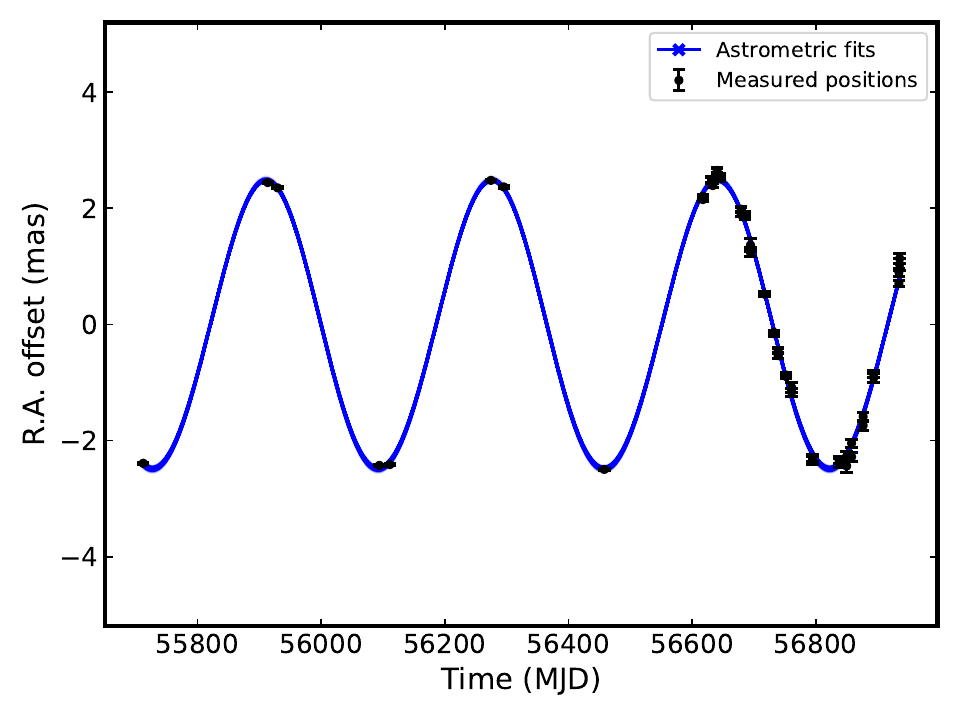}
  \includegraphics[width=0.49\linewidth]{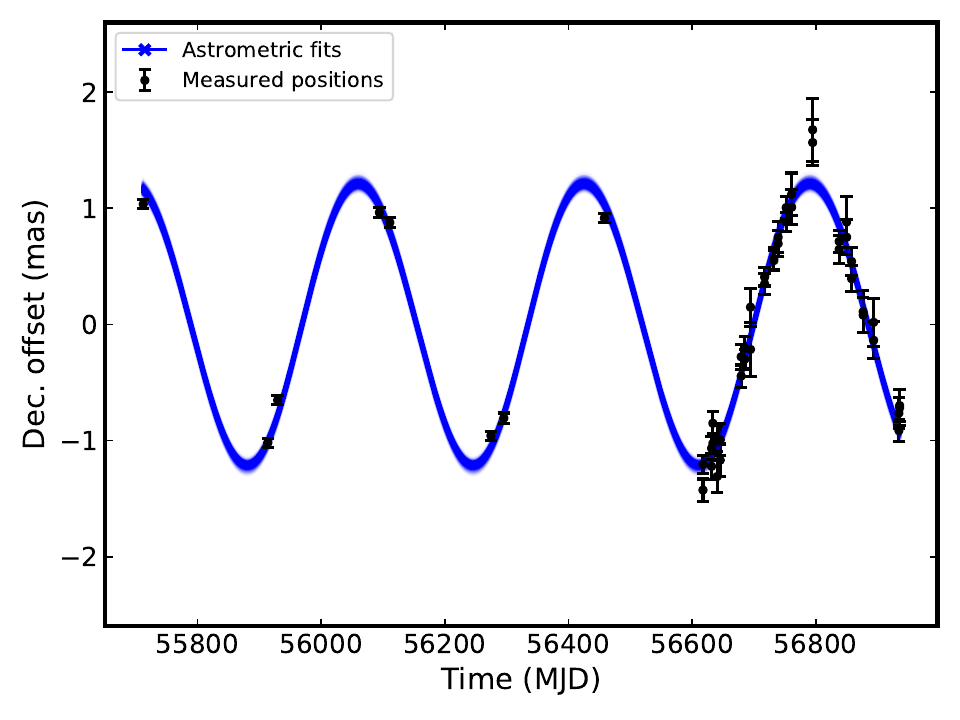}
  \includegraphics[width=0.49\linewidth]{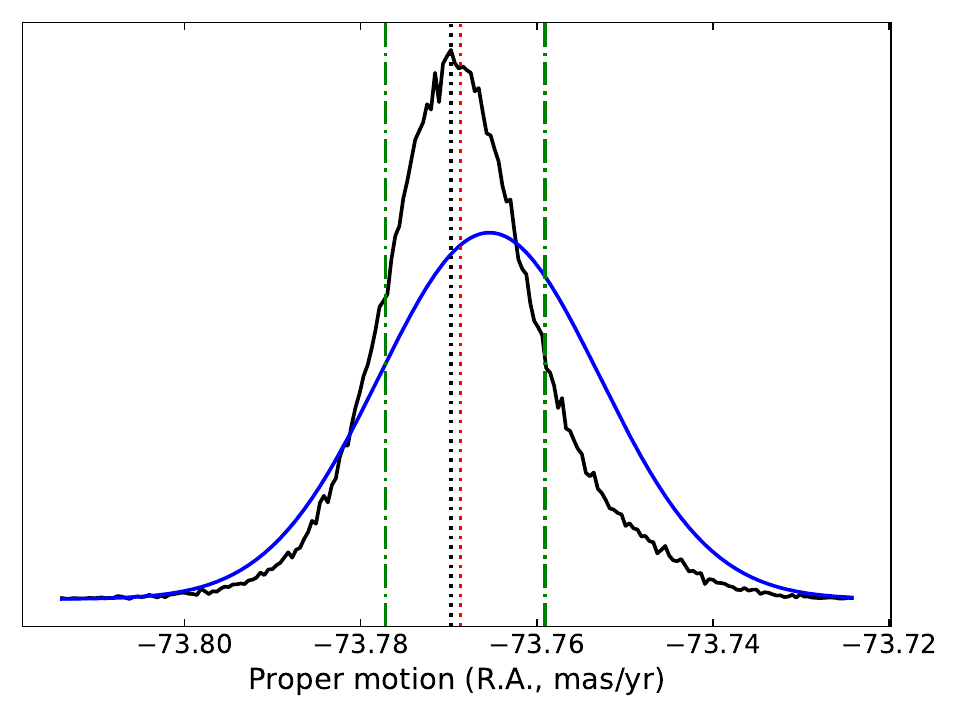}
  \includegraphics[width=0.49\linewidth]{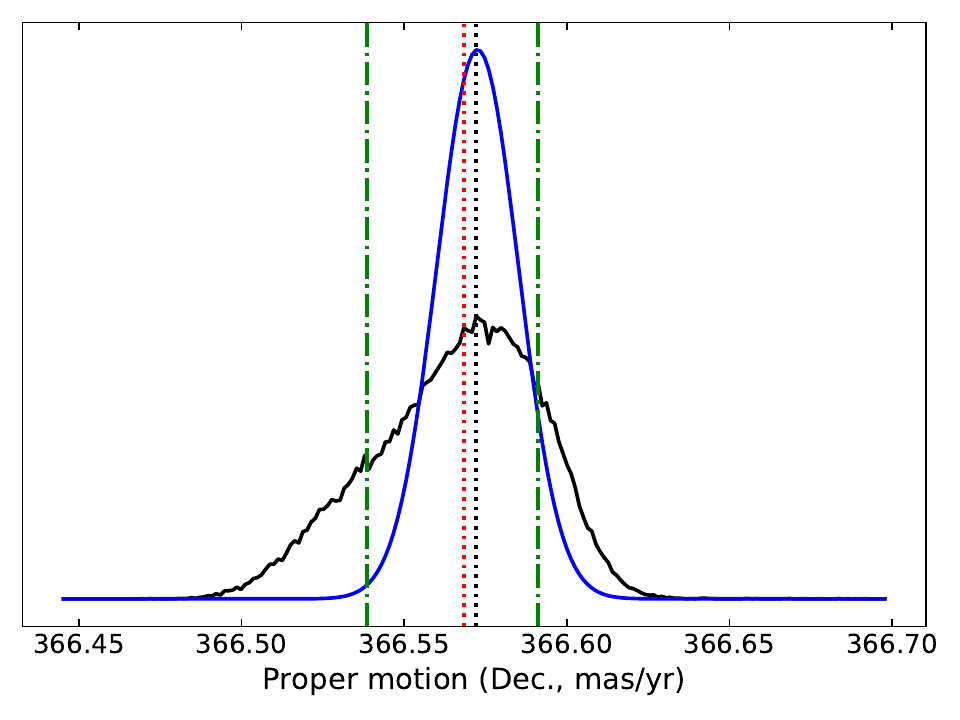}
  \includegraphics[width=0.49\linewidth]{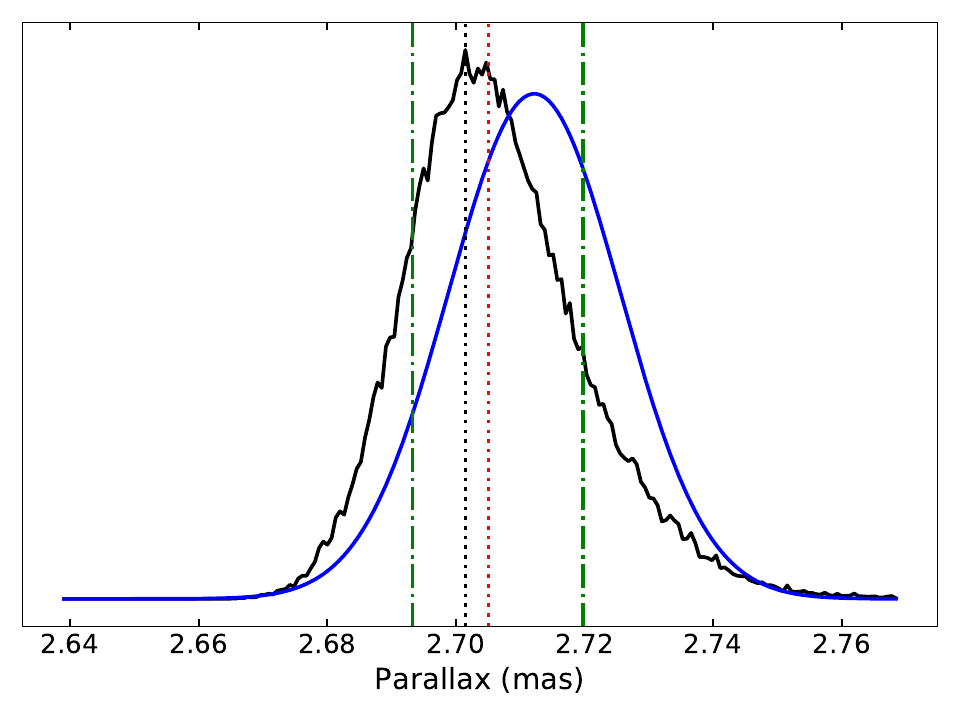}  
  \caption{Bootstrap fitting estimates for B1133+16. In the top panel plots, the first eight data points (from left) are from BD152 observations, and the rest of the data points are from BD174 observations. For details, see the caption of Figure \ref{fig:boot_results_J0332}.}
  \label{fig:boot_results_J1136}
\end{figure*}

Here, a large number of bootstrap sample sets (100000; each of size equal to observation epochs, N$_e$) are generated by randomly choosing any of the epoch entries with replacement from the given input data set, where some of the epoch entries from the original set might repeat and/or be absent. If N$_e$ is small, then a bootstrap sample may have fewer than three unique epochs, which are required to solve for the aforementioned five astrometric parameters. We have set the required minimum unique epochs to five to reduce the biasing in the fit due to the chance selection of unique epochs at one side of the parallax ellipse, and all with a small parallax contribution to estimate the underlying parallax swing. If a bootstrap sample has fewer than the minimum unique epochs, then the sample is not considered in the bootstrap fitting. For each valid bootstrap sample set, astrometric parameters are inferred using \texttt{PMPAR} software, and a probability density distribution is examined for each parameter from which the median value of the parameter and the $84^{th}$ and $16^{th}$ percentile values ($1\sigma$ uncertainty) are estimated. The bootstrap fitting estimates for both pulsars are shown in Figures \ref{fig:boot_results_J0332} and \ref{fig:boot_results_J1136} and tabulated in Table \ref{table:ast_results_56000}. For comparison, we also include the astrometric estimates for these pulsars from \cite{Deller_2019}. However, they reported the most probable values of the parameters and used the most compact 68\% confidence interval around the parameter value to estimate the $1\sigma$ error bars, which is less robust, particularly for asymmetric distributions. Our bootstrap parameter estimates from the combined data set (BD152+BD174) are consistent with those of \cite{Deller_2019} within the reported uncertainties, but with a significant improvement in precision (Table \ref{table:ast_results_56000}). As expected, the enhancement is more pronounced in proper motion than in parallax since the proper motion precision grows with observation time as $t^{3/2}$, whereas parallax precision grows with $t^{1/2}$ (assuming all observations are of equal sensitivity and equally spaced in time).

The BD174 data set exhibits comparatively larger error bars (Figures \ref{fig:boot_results_J0332} and \ref{fig:boot_results_J1136}) due to a couple of reasons: a) higher thermal noise resulting from less than half the target scan time compared to BD152, b) observations taken during the solar maxima; the typical mean Vertical TEC (VTEC) value during BD174 observations $\sim25$\,TECU ($1\,{\rm TECU} = 10^{16}\,{\rm electrons\, m^{-2}}$), which is considerably higher than the $\sim17$\,TECU seen during BD152 observations (estimated by randomly sampling 5--10 epochs from each observation series), c) bigger beam size due to missing antennas (sometimes) and low-frequency observation in the case of 1.4\,GHz data set.

3) \textbf{Bayesian approach:}\label{sec:4.3}
We have also inferred astrometric parameters based on Bayesian statistics using \texttt{STERNE}\footnote{\url{https://github.com/dingswin/sterne}} package \citep{Ding_2021, 10.1093/mnras/stac3725}. This tool is specifically designed for VLBI astrometry and allows estimation of additional parameters, such as orbital parameters in the case of a binary system and $\eta_{\text{EFAC}}$, a scale factor to account for the underestimation or overestimation of systematic uncertainties. The $\eta_{\text{EFAC}}$ factor is assumed to be one in the least-squares fitting and bootstrap approach. The total error at the $i^{th}$ epoch is
\begin{equation}
    \sigma^i({\eta_{\text{EFAC}}}) = \sqrt{(\sigma_R^i)^2 + (\eta_{\text{EFAC}} \cdot \sigma_{\text{sys}}^i)^2},
    \label{eq:random_sys_quad}
\end{equation}
where $\sigma_R$ and $\sigma_{\text{sys}}$ are the random and systematic errors. 

\begin{figure*}
    \centering
    \includegraphics[width=1\linewidth]{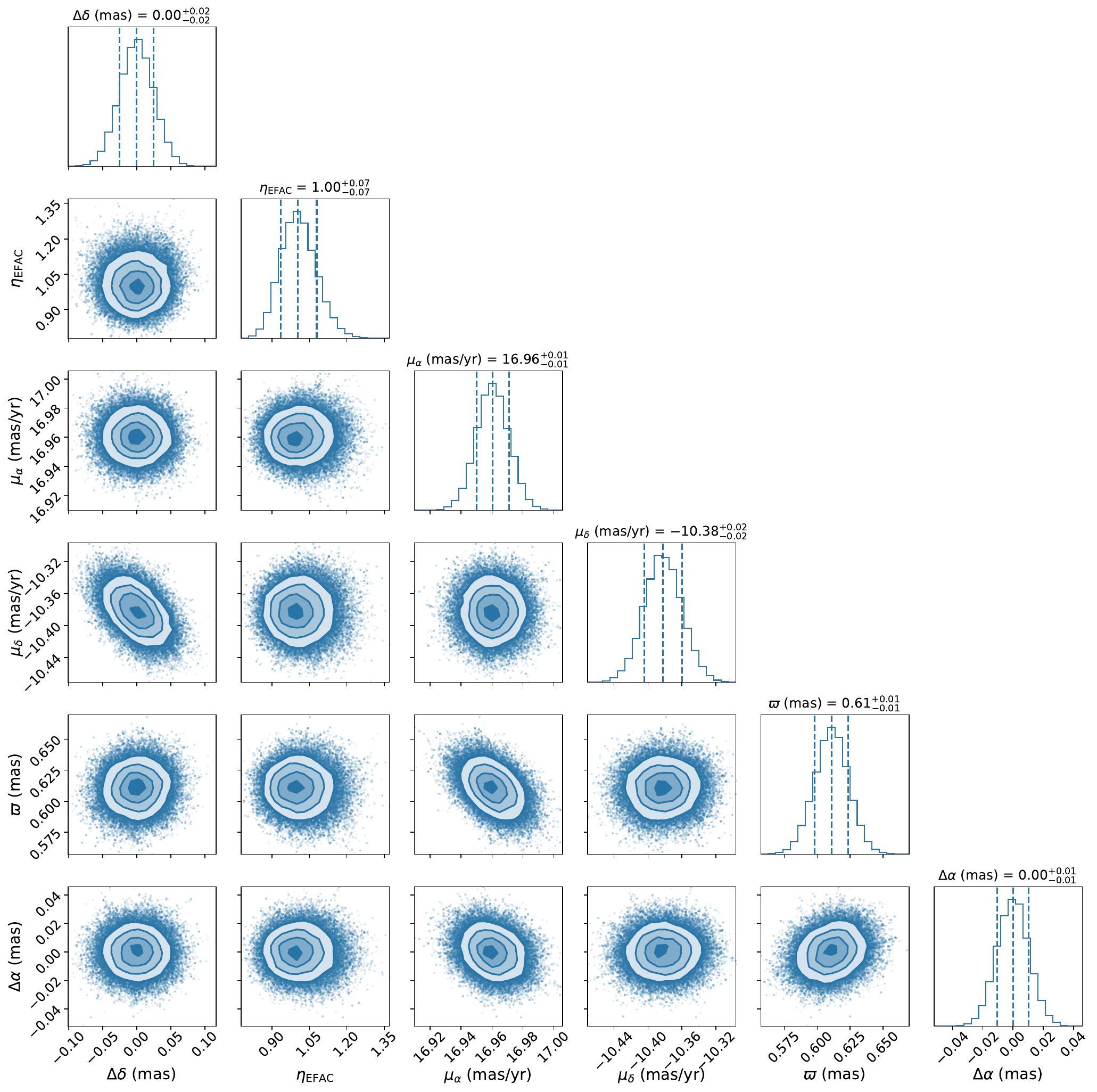} 
    \caption{Error ellipses and marginalised histograms are plotted for $\eta_{\rm EFAC}$ and five astrometric parameters of B0329+54 using the Bayesian approach. The contours show the $1\sigma$, $2\sigma$, and $3\sigma$ confidence intervals, and the position offset ($\Delta\alpha$ and $\Delta\delta$) is with respect to the reference epoch position (MJD: 56300), while dashed lines in histograms mark the median value of astrometric parameters and $1\sigma$ deviations.}
    \label{fig:bayesian_J0332}
\end{figure*}

\begin{figure*}
    \centering
    \includegraphics[width=1\linewidth]{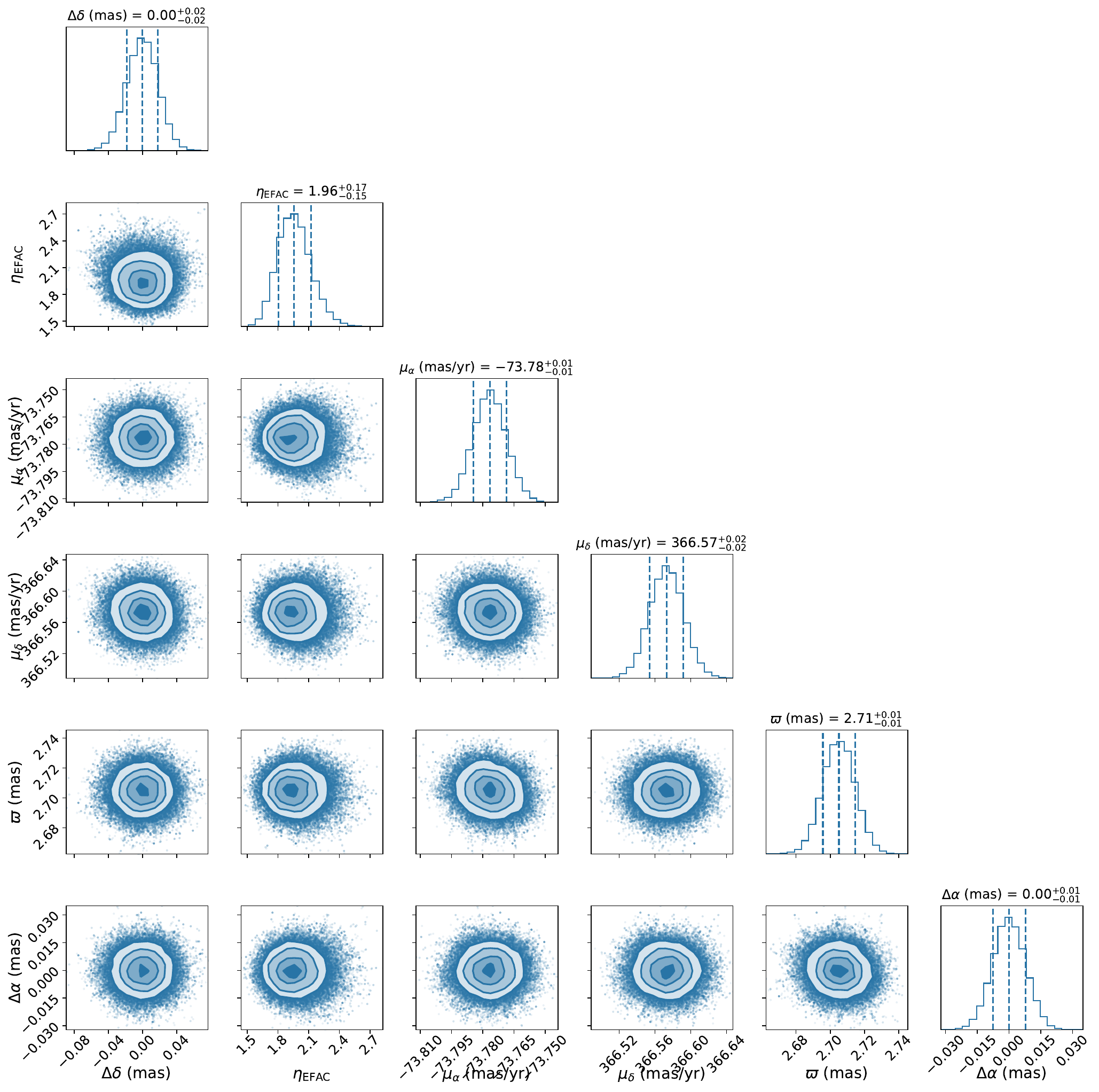} 
    \caption{Error ellipses and marginalised histograms are plotted for $\eta_{\rm EFAC}$ and five astrometric parameters of B1133+16 using the Bayesian approach. For details, see the caption of Figure \ref{fig:bayesian_J0332}.}
    \label{fig:bayesian_J1136}
\end{figure*}

The priors for astrometric parameters are assumed uniformly distributed, $\mathcal{U}(X^\text{DF} - 20 \widetilde{\sigma}^\text{DF}_X, X^\text{DF} + 20 \widetilde{\sigma}^\text{DF}_X)$, where $X^\text{DF}$ and $\widetilde{\sigma}^\text{DF}_X$ stand for astrometric parameters and associated scaled uncertainties ($\widetilde{\sigma}^\text{DF}_X = \sqrt{\chi^2}\sigma^\text{DF}_X$), taken from the least-squares fitting, and $\mathcal{U}(0, 15)$ for $\eta_{\text{EFAC}}$. The uniform distribution avoids biasing, as we have used \texttt{PMPAR} results obtained from the same data to set the priors. Figures \ref{fig:bayesian_J0332} and \ref{fig:bayesian_J1136} illustrate the inferred astrometric parameters for B0329+54 and B1133+16 using the Bayesian approach alongside $\eta_{\rm EFAC}$, and the median value of astrometric parameters along with their associated uncertainties are tabulated in Table \ref{table:ast_results_56000}. The $\eta_{\text{EFAC}}$ value of unity and two for B0329+54 and B1133+16, respectively, indicates that the initially estimated systematic uncertainties are underestimated for the latter.

All astrometric parameters obtained using the three mentioned statistical techniques are consistent within the error bars. However, the bootstrap and Bayesian methods perform better, as they rely less on prior knowledge of uncertainties in the input data. In particular, the Bayesian approach proves more robust when reliable prior information is available for the fitting parameters. Therefore, we will only consider Bayesian estimates for the remainder of the discussion.

For each pulsar, the proper motion and parallax estimates obtained from 1.4\,GHz and 1.6\,GHz data sets are consistent within their error bars, except for the parallax of B1133+16, which deviates at the $1.5 \sigma$ level (Table \ref{table:ast_results_56300}). As expected, the 1.6\,GHz data set provides better astrometric precision, and the combination of 1.4\,GHz and 1.6\,GHz further enhances the precision of astrometric parameters by approximately 1.2 times. Similarly, incorporating the BD174 data set alongside BD152 improves astrometric precision by increasing the number of observations and, more importantly, extending the observation time span. Comparing the Bayesian estimates of the inferred parameters from the BD152-only data set (Table \ref{table:ast_results_56000}, reference epoch MJD 56000) with those from the combined BD152+BD174 data set (Table \ref{table:ast_results_56300}, reference epoch MJD 56300), the proper motion and parallax estimates for both pulsars remain consistent within the respective error bars. Both analyses used the traditional mapping function. However, the combined data set yields almost threefold improvement in proper motion precision, and parallax precision improves by nearly a factor of two.
\begin{table*}[htbp]
    \centering
    \resizebox{\textwidth}{!}{%
    \begin{threeparttable}
    \begin{tabular}{ccccc}
    \toprule
    Method & $\mu_{\alpha} = \dot{\alpha} \cos\delta$ & $\mu_{\delta} = \dot{\delta}$ & $\boldsymbol{\varpi}$ & $\boldsymbol{D}$ \\
    & (mas\,yr$^{-1}$) & (mas\,yr$^{-1}$) & (mas) & (pc) \\
    \midrule
    B0329+54, fitting with only BD152 data set \\
    Least Squares & 16.961 $\pm$ 0.013 & -10.377 $\pm$ 0.029 & 0.607 $\pm$ 0.011 & $1650^{+30}_{-30}$ \\
    Bootstrap Fitting & $16.959^{+0.040}_{-0.029}$ & $-10.376^{+0.065}_{-0.048}$ & $0.604^{+0.030}_{-0.030}$ & $1660^{+90}_{-80}$ \\
    Bayesian Inference & $16.959^{+0.030}_{-0.029}$ & $-10.377^{+0.065}_{-0.064}$ & $0.608^{+0.025}_{-0.026}$ & $1640^{+70}_{-70}$ \\
    \cite{Deller_2019} (Bootstrap) & $16.969^{+0.027}_{-0.029}$ & $-10.379^{+0.058}_{-0.036}$ & $0.595^{+0.020}_{-0.025}$ & $1680^{+70}_{-70}$ \\
    \midrule
    B0329+54, fitting with both (BD152+BD174) data sets \\
    Least Squares & 16.963 $\pm$ 0.008 & -10.384 $\pm$ 0.016 & 0.611 $\pm$ 0.009 & $1640^{+30}_{-20}$ \\
    Bootstrap Fitting & $16.963^{+0.015}_{-0.015}$ & $-10.384^{+0.024}_{-0.020}$ & $0.613^{+0.023}_{-0.019}$ & $1630^{+50}_{-60}$ \\
    Bayesian Inference & $16.964^{+0.011}_{-0.011}$ & $-10.383^{+0.022}_{-0.022}$ & $0.610^{+0.013}_{-0.013}$ & $1640^{+40}_{-30}$ \\
    \midrule
    B1133+16, fitting with only BD152 data set \\
    Least Squares & -73.771 $\pm$ 0.006 & 366.567 $\pm$ 0.015 & 2.696 $\pm$ 0.004 & 370.9$^{+0.6}_{-0.6}$ \\
    Bootstrap Fitting & $-73.770^{+0.026}_{-0.012}$ & $366.567^{+0.065}_{-0.084}$ & $2.696^{+0.016}_{-0.014}$ & $371^{+2}_{-2}$ \\
    Bayesian Inference & $-73.765^{+0.024}_{-0.025}$ & $366.564^{+0.062}_{-0.064}$ & $2.697^{+0.018}_{-0.018}$ & $368^{+2}_{-2}$ \\
    \cite{Deller_2019} (Bootstrap) & $-73.785^{+0.031}_{-0.010}$ & $366.569^{+0.072}_{-0.055}$ & $2.687^{+0.018}_{-0.016}$ & $372^{+2}_{-2}$ \\
    \midrule
    B1133+16, fitting with both (BD152+BD174) data sets \\
    Least Squares & -73.769 $\pm$ 0.003 & 366.568 $\pm$ 0.008 & 2.703 $\pm$ 0.004 & $370.0^{+0.5}_{-0.5}$ \\ 
    Bootstrap Fitting & $-73.769^{+0.010}_{-0.008}$ & $366.568^{+0.023}_{-0.029}$ & $2.705^{+0.015}_{-0.012}$ & $370^{+2}_{-2}$ \\
    Bayesian Inference & $-73.776^{+0.008}_{-0.008}$ & $366.573^{+0.019}_{-0.020}$ & $2.704^{+0.010}_{-0.009}$ & $370^{+1}_{-1}$ \\
    \bottomrule
    \end{tabular}%
    \end{threeparttable}
    }
    \caption{Astrometric parameters obtained using all three statistical approaches. For comparison, \cite{Deller_2019} estimates for these pulsars are also included. To facilitate direct comparison, the position of each pulsar at the reference epoch (MJD: 56000) has been aligned with that of \cite{Deller_2019} and is therefore not listed in this table.}
    \label{table:ast_results_56000}
\end{table*}

\section{Effect of TEC mapping function on astrometric parameters}\label{sec:5}
The TECOR task performs a priori ionospheric delay corrections by computing the excess path delay using global TEC maps. For both data sets, the ionospheric dispersive delays are estimated using TEC maps provided by the International Global Navigation Satellite System (GNSS) Service (centre code: \texttt{IGSG}). These maps represent the VTEC distribution on a global grid with a spatial and temporal resolution of $5^\circ\, \times\, 2.5^\circ$ (${\rm longitude}\, \times\, {\rm latitude}$) and 2\,h, respectively. Co-located GNSS receivers at VLBI stations have also been characterised \citep{skeens2023}, and may be useful in future for deriving site-specific TEC measurements and enabling real-time ionospheric corrections.

The TECOR task utilises a mapping function to convert the VTEC to Slant TEC (STEC). The traditional mapping function is the so-called single-layer model, $M(e)\, =\, 1/\cos({\rm ZA})$, where ZA is the Zenith angle at the ionosphere puncture point. Recently, \cite{Petrov_2023} has studied a so-called thin-shell ionospheric mapping function, suggested by \cite{1999GGAS...59.....S},
\begin{equation}
M(e) = k \dfrac{1}{\sqrt{1 - \left(\dfrac{R_{\oplus}}{R_{\oplus} + H_i + \Delta H}\right)^2 \cos^2\alpha e_{gc}}},
\label{eq:petrov23}
\end{equation}
here, $R_{\oplus}$ denotes the Earth's radius, $e_{gc}$ is the geocentric elevation angle with respect to the radius vector between the geocentre and the station, and $H_i$ is the height of the ionosphere. \cite{Petrov_2023} has recommended values for scaling factor ($k$), thickness of the ionospheric layer ($\Delta H$), and fudge factor ($\alpha$) to 0.85, 56.7\,km, and 0.9782, respectively. The modified (Petrov23) mapping function aims to enhance the accuracy of ionospheric corrections based on global TEC maps, and has been implemented in the TECOR task of \texttt{AIPS} since 31DEC23. We emphasise that these global ionospheric maps only capture the bulk/smooth component of the ionosphere, due to their limited time and spatial resolution, and small-scale or rapidly varying ionospheric disturbances will remain uncorrected regardless of the TEC mapping function used. 

\begin{figure*}[htbp]
    \centering
    \includegraphics[width=0.49\linewidth]{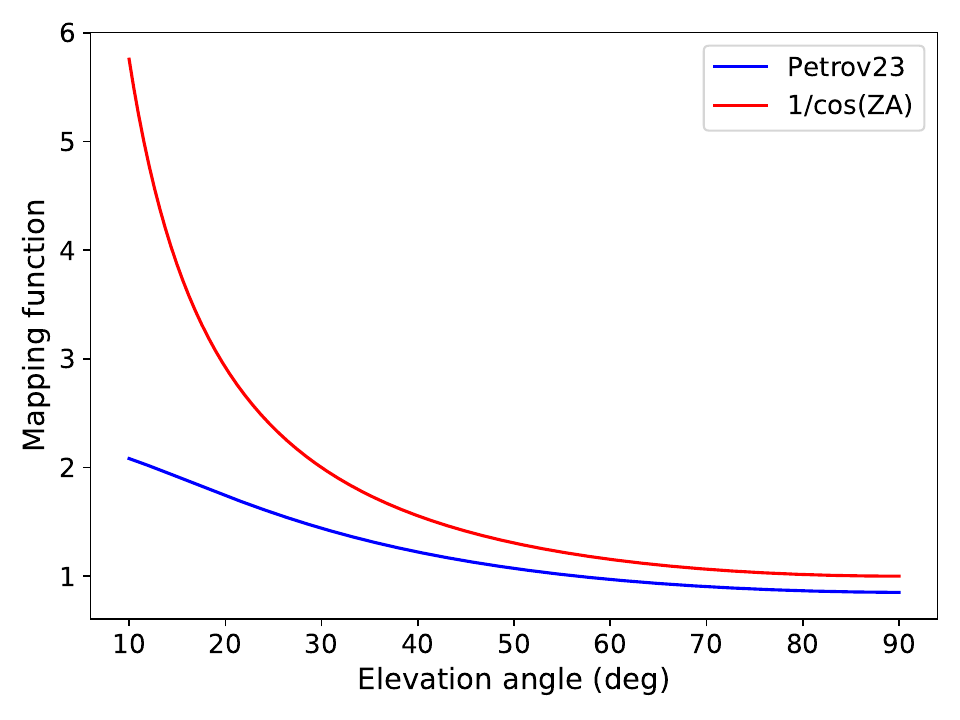}
    \includegraphics[width=0.49\linewidth]{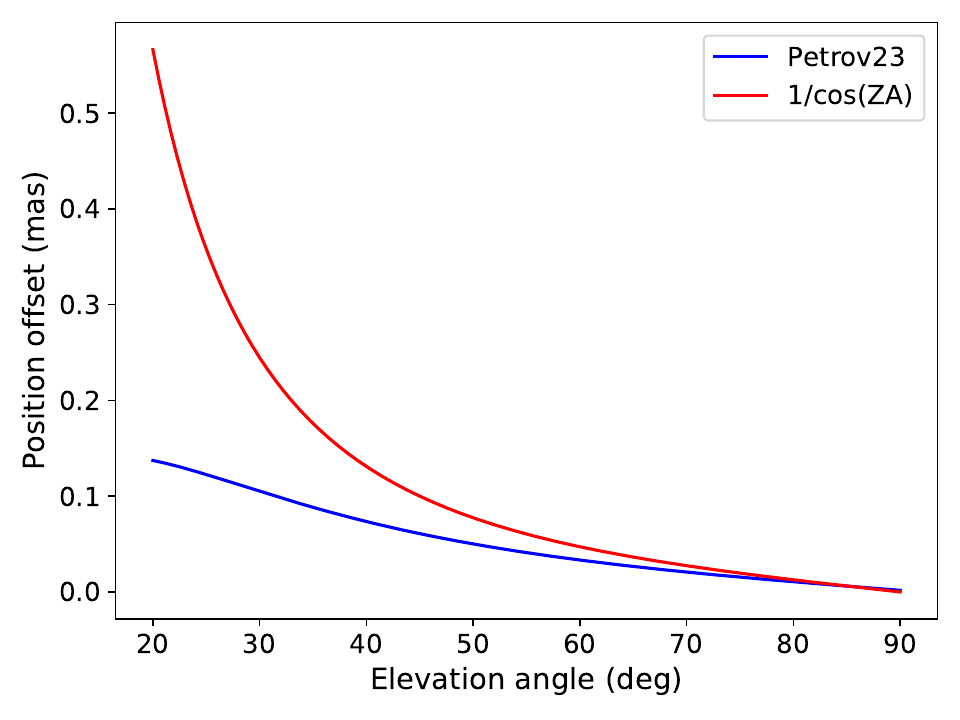}
    \caption{The left plot shows the difference between the traditional and Petrov23 mapping functions (equivalent to STEC/VTEC), while the right plot illustrates the position offset caused by residual dispersive delay at 1.6\,GHz when the delay is computed using the respective mapping function. A nominal TEC value of 10\,TECU is assumed, and the IBC is assumed at six arcminutes above the target.}
    \label{fig:map_fns}
\end{figure*}

These two mapping functions exhibit an offset in STEC prediction that becomes increasingly significant at lower elevations. Even at an elevation of $90^\circ$, the Petrov23 mapping function predicts a dispersive delay that is approximately 20\% lower than that of the traditional mapping function for a given VTEC value (Figure \ref{fig:map_fns}). These differences affect the inferred source position and can introduce noticeable position shifts even across the small angular separations typical for differential astrometry. For example, consider a field in which the target is separated from a calibrator source by six arcminutes in declination, observed at an elevation of $35^\circ$, at a frequency of 1.6\,GHz, and with a maximum baseline length of 8000\,km. When ionospheric delays are corrected using TEC maps (assuming a nominal TEC value of 10\,TECU) utilising each of the mapping functions, the resulting position shift can be as large as $\sim0.1$\,mas (Figure \ref{fig:map_fns}).

To examine the impact of TEC mapping functions on astrometry, we have processed the data twice, using each mapping function and compared the resulting position-time series. For each pulsar, the median of the difference of position-time series obtained using traditional and Petrov23 mapping functions is shown in Table \ref{table:median_pos_diff}. The quantity determining the shift in the target position, assuming the shift is only due to the switching between mapping functions, is the difference in dispersive delays (computed by TECOR) between the IBC and target directions when using the traditional and Petrov23 mapping functions. The obtained position-time series difference is within the expectation from the position difference shown in Figure \ref{fig:map_fns} at higher elevations.
\begin{table}[htbp]
    \centering
    \resizebox{\textwidth}{!}{%
    \begin{tabular}{cccc}
    \toprule
    Pulsar & $\Delta {\rm P}_{1.4}$ & $\Delta {\rm P}_{1.6}$ & $\Delta {\rm P}_{\rm BD152}$ \\
     & (mas) & (mas) & (mas) \\
    \midrule  
    B0329+54 & (-0.065, 0.039) & (-0.054, 0.037) & (-0.018, 0.018) \\
    \addlinespace
    B1133+16 & (-0.004, 0.002) & (-0.003, 0.002) & (-0.010, 0.007) \\
    \bottomrule
    \end{tabular}%
    }
    \caption{Median of position-time series differences ($\Delta\alpha$, $\Delta\delta$) obtained using each mapping function (Figures \ref{fig:pos_diff_J0332} and \ref{fig:pos_diff_J1136}). The second and third columns correspond to results from the BD174 data set.}
    \label{table:median_pos_diff}
\end{table}

The astrometric parameters obtained using each mapping function are shown in Table \ref{table:ast_results_56300}. The difference in the fitted astrometric parameters is within the respective error bars. However, it is not apparent that the Petrov23 mapping function is definitely an improvement, by examining the precision of the astrometric parameters and the reduced chi-square values (in least-squares fitting) for different combinations of data sets. Although the reduced chi-square values are not appreciably better or worse for the full BD152+BD174 data in either case, nor are the BD174-only results consistently closer to the truth (where we assume that BD152+BD174 gives a reasonable approximation of truth, at least compared to BD174 alone) with one mapping function compared to the other. We conclude that significant model uncertainty remains in the application of ionospheric corrections based on global ionosphere models. In any case, ionospheric variations on small spatial and temporal scales that are not captured by global TEC maps (regardless of the mapping function used) may be more significant for differential astrometric observations than the difference between the mapping function used in global TEC maps, at least at small angular separations typical for in-beam calibrator observations.
\begin{table*}[htbp]
    \centering
    \resizebox{\textwidth}{!}{%
    \begin{threeparttable}
    \begin{tabular}{ccccccc}
    \toprule
    Project/band & $\Delta\alpha_\text{J2000}$\tnote{1} & $\Delta\delta_\text{J2000}$\tnote{1} & $\mu_{\alpha}=\dot{\alpha}\cos\delta$ & $\mu_{\delta}=\dot{\delta}$ & $\boldsymbol{\varpi}$ & $\boldsymbol{D}$ \\
    & (mas) & (mas) & (mas\,yr$^{-1}$) & (mas\,yr$^{-1}$) & (mas) & (pc) \\
    \midrule
    \addlinespace 
    B0329+54, traditional mapping function\\
    \addlinespace
    BD174 (1.4\,GHz) & 1.72(3) & 0.9(4) & 17.063$^{+0.185}_{-0.182}$ & -10.831$^{+0.261}_{-0.258}$ & 0.611$^{+0.064}_{-0.064}$ & $1640^{+190}_{-160}$ \\
    \addlinespace   
    BD174 (1.6\,GHz) & -0.76(2) & -0.4(2) & 17.078$^{+0.111}_{-0.114}$ & -10.546$^{+0.166}_{-0.160}$ & 0.595$^{+0.040}_{-0.041}$ & $1680^{+120}_{-110}$ \\
    \addlinespace 
    BD174+BD152 & -0.960(1) & -0.43(3) & $16.960^{+0.011}_{-0.010}$ & $-10.382^{+0.022}_{-0.022}$ & $0.611^{+0.013}_{-0.013}$ & $1640^{+30}_{-30}$ \\
    \addlinespace 
    B0329+54, Petrov23 mapping function\\
    \addlinespace
    BD174 (1.4\,GHz) & 1.93(3) & 0.7(4) & 17.189$^{+0.192}_{-0.183}$ & -10.906$^{+0.264}_{-0.266}$ & 0.545$^{+0.064}_{-0.067}$ & $1830^{+240}_{-200}$ \\
    \addlinespace   
    BD174 (1.6\,GHz) & -0.61(2) & -0.5(2) & 17.167$^{+0.115}_{-0.116}$ & -10.594$^{+0.163}_{-0.161}$ & 0.552$^{+0.041}_{-0.041}$ & $1810^{+140}_{-130}$ \\
    \addlinespace  
    BD174+BD152 & -0.988(1) & -0.45(3) & $16.961^{+0.012}_{-0.011}$ & $-10.386^{+0.023}_{-0.023}$ & $0.598^{+0.013}_{-0.014}$ & $1670^{+40}_{-40}$ \\
    \addlinespace 
    B1133+16, traditional mapping function\\
    \addlinespace
    BD174 (1.4\,GHz) & 1.21(1) & 1.3(2) & -73.549$^{+0.139}_{-0.142}$ & 366.671$^{+0.187}_{-0.189}$ & 2.842$^{+0.054}_{-0.055}$ & 352$^{+7}_{-7}$ \\
    \addlinespace   
    BD174 (1.6\,GHz) & -0.565(8) & -0.6(2) & -73.717$^{+0.095}_{-0.091}$ & 366.499$^{+0.133}_{-0.131}$ & 2.740$^{+0.038}_{-0.037}$ & 365$^{+5}_{-5}$ \\
    \addlinespace 
    BD174+BD152 & -0.6490(5) & -0.72(2) & $-73.777^{+0.008}_{-0.008}$ & $366.573^{+0.019}_{-0.019}$ & $2.705^{+0.009}_{-0.009}$ & $370^{+1}_{-1}$ \\
    \addlinespace 
    B1133+16, Petrov23 mapping function\\
    \addlinespace
    BD174 (1.4\,GHz) & 1.25(1) & 1.5(2) & $-73.521^{+0.144}_{-0.147}$ & $366.710^{+0.185}_{-0.187}$ & $2.847^{+0.055}_{-0.056}$ & 351$^{+7}_{-7}$ \\
    \addlinespace   
    BD174 (1.6\,GHz) & -0.589(9) & -0.5(2) & -73.733$^{+0.101}_{-0.099}$ & 366.528$^{+0.139}_{-0.139}$ & 2.733$^{+0.039}_{-0.040}$ & 366$^{+5}_{-5}$ \\
    \addlinespace  
    BD174+BD152 & 1.7455(6) & -0.53(2) & $-73.779^{+0.010}_{-0.009}$ & $366.578^{+0.022}_{-0.022}$ & $2.703^{+0.011}_{-0.011}$ & $370^{+1}_{-1}$ \\
    \bottomrule
    \end{tabular}%
    \begin{tablenotes}
    \item[1] The position offsets ($\Delta\alpha_\text{J2000}$, $\Delta\delta_\text{J2000}$) are with respect to the mean reference position of B0329+54 and B1133+16, which are, R.A.= 03:32:59.4109883, Dec.= 54:34:43.31937 and R.A.= 11:36:03.1154509, Dec.= 15:51:14.48331, respectively. The reference positions are with respect to the IBC at a reference epoch, MJD: 56300, and associated error bars do not include the position uncertainties in the PRC and IBC and other systematic errors.
    \end{tablenotes}
    \end{threeparttable}
    }
    \caption{Bayesian estimates of astrometric parameters for B0329+54 and B1133+16 utilising each TEC mapping function to compute the ionospheric dispersive delays and different combinations of data sets included in the fitting.}
    \label{table:ast_results_56300}
\end{table*}

\begin{figure*}[htbp]
    \centering
    \includegraphics[width=0.49\linewidth]{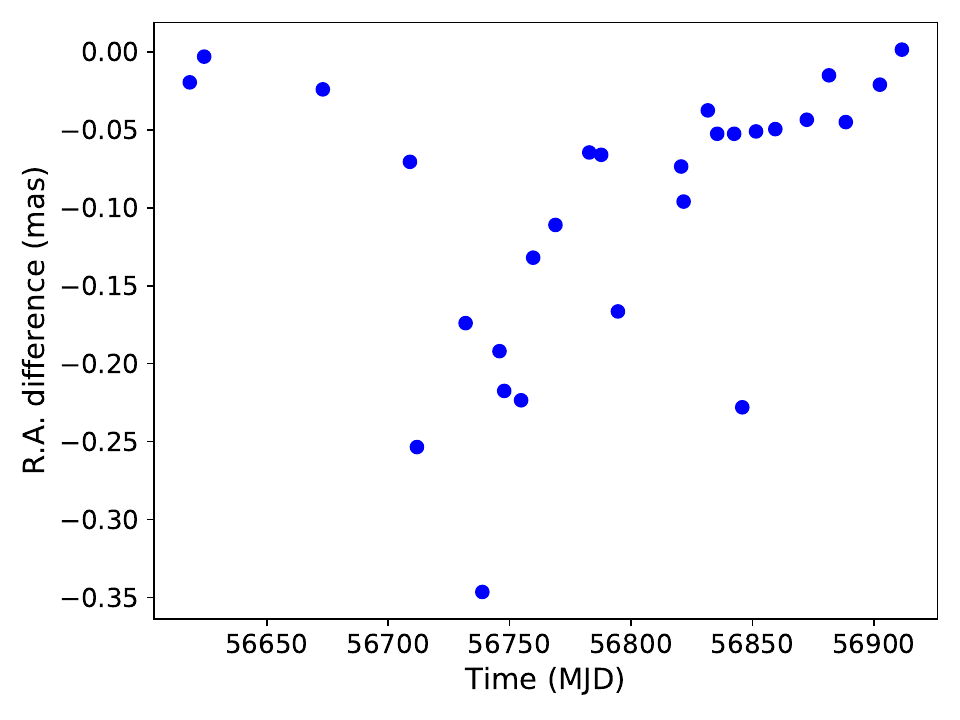}
    \includegraphics[width=0.49\linewidth]{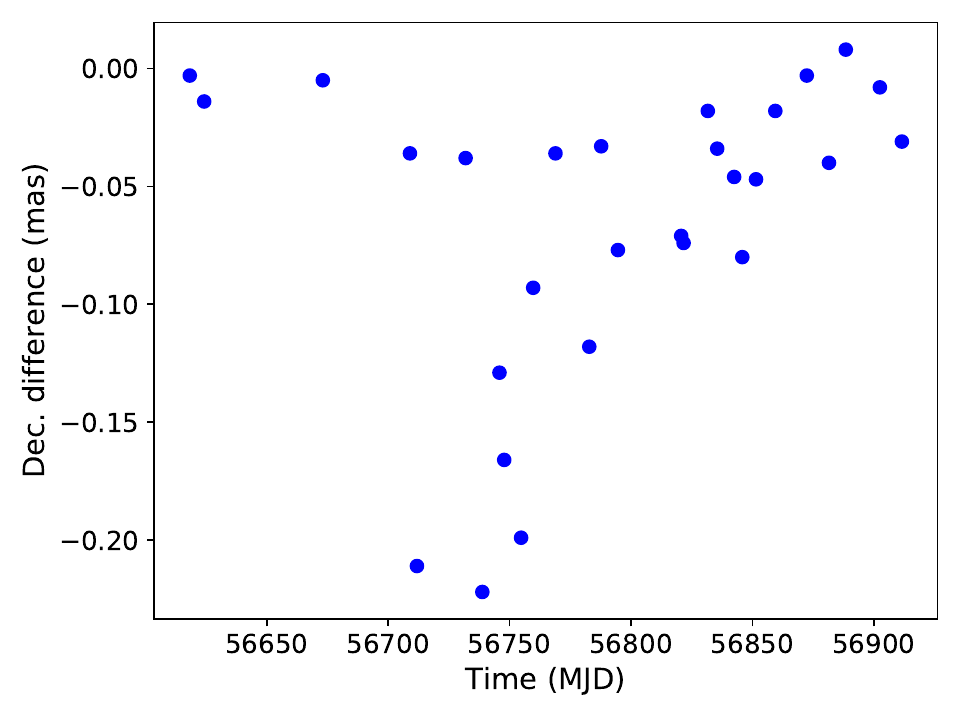}
    \includegraphics[width=0.49\linewidth]{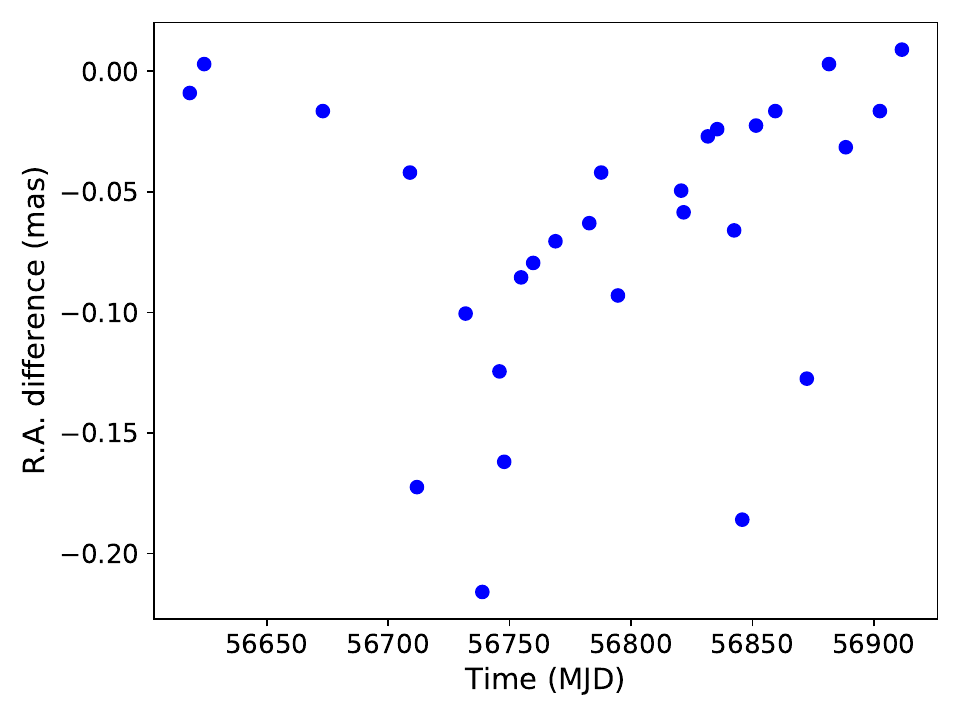}
    \includegraphics[width=0.49\linewidth]{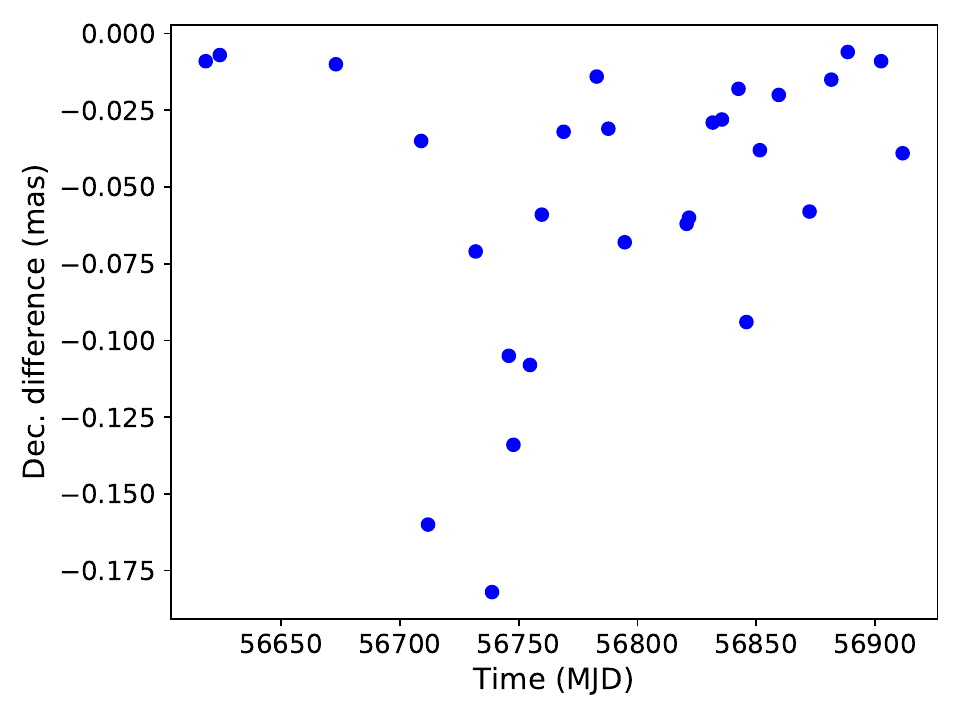}
    \includegraphics[width=0.49\linewidth]{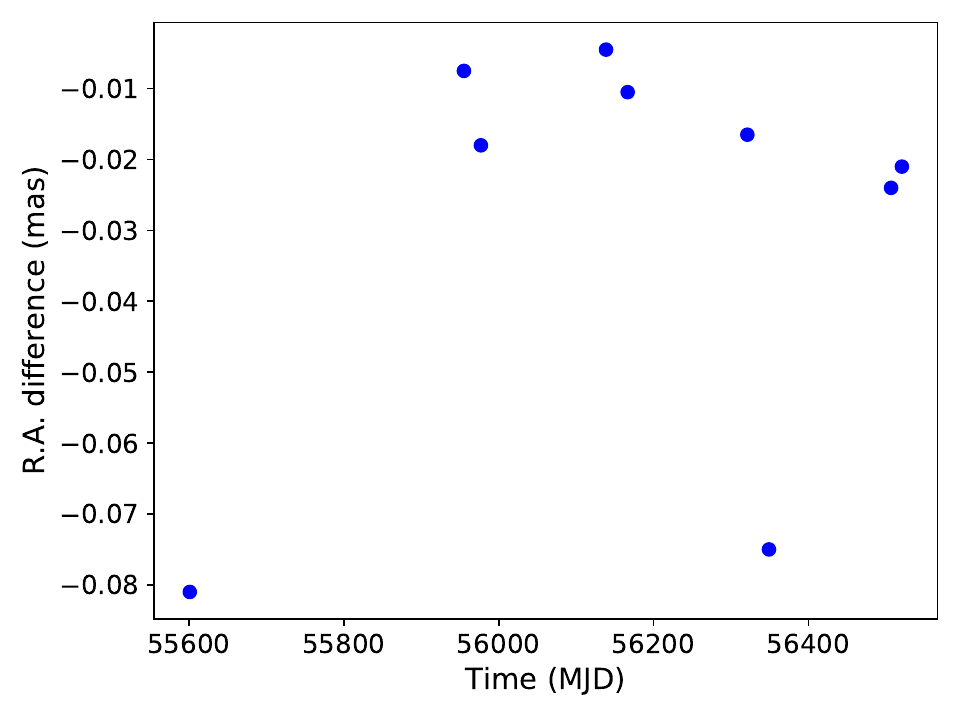}
    \includegraphics[width=0.49\linewidth]{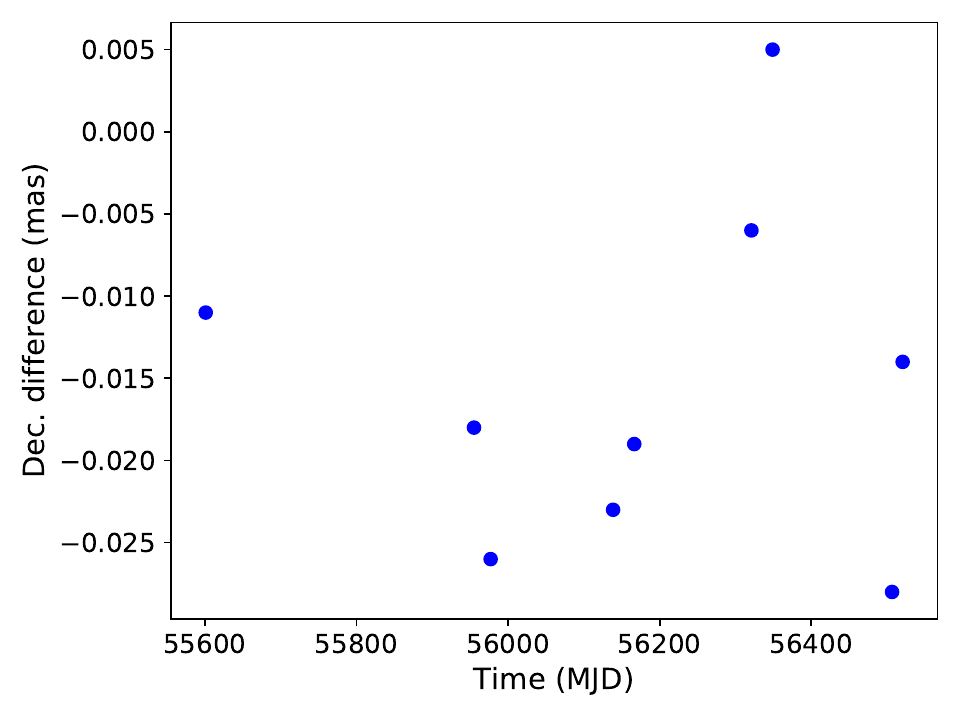}  
    \caption{The plots show the B0329+54 position-time series difference (P$_{\rm trad}$ - P$_{\rm Petrov23}$), when the data have been processed using the traditional and Petrov23 mapping functions. The position error bars are not included, since they are comparable to the difference or larger for some epochs. The top, middle, and bottom panels of the figure present the BD174 (1.4\,GHz), BD174 (1.6\,GHz), and BD152 data sets. The position differences at 1.6\,GHz are smaller compared to 1.4\,GHz, indicating the smaller residual dispersive delays at higher frequencies.}
    \label{fig:pos_diff_J0332}
\end{figure*}

\begin{figure*}[htbp]
    \centering
    \includegraphics[width=0.49\linewidth]{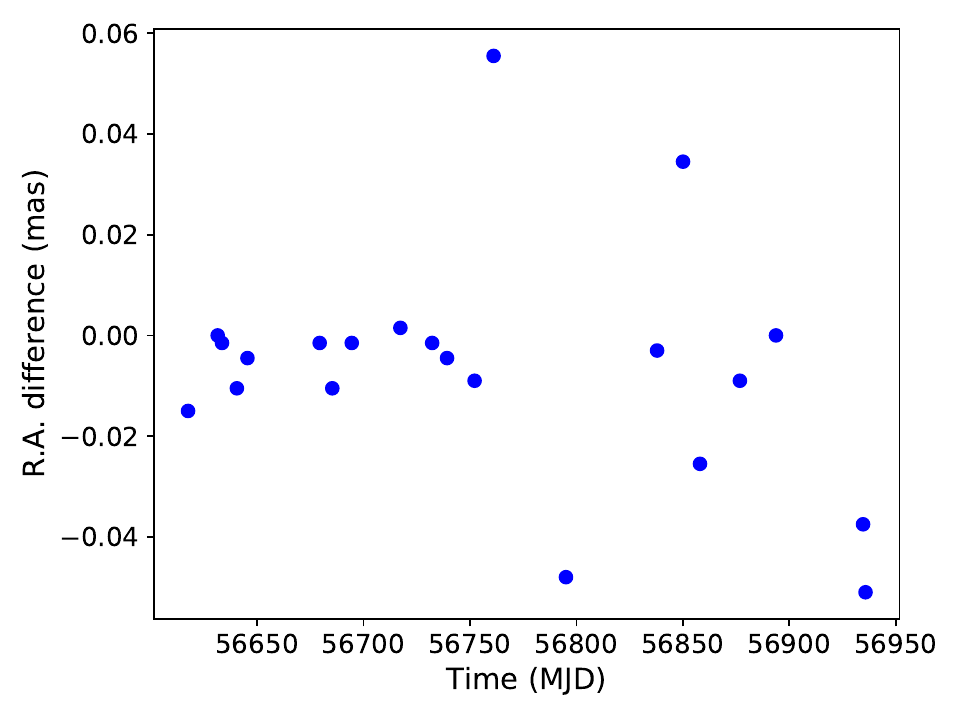}
    \includegraphics[width=0.49\linewidth]{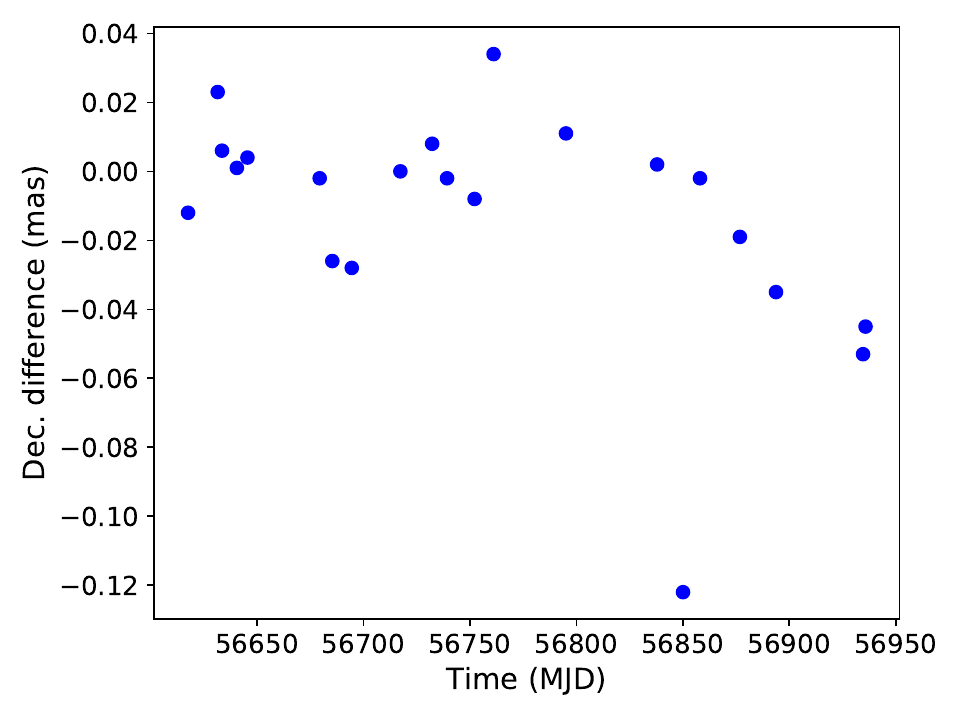}
    \includegraphics[width=0.49\linewidth]{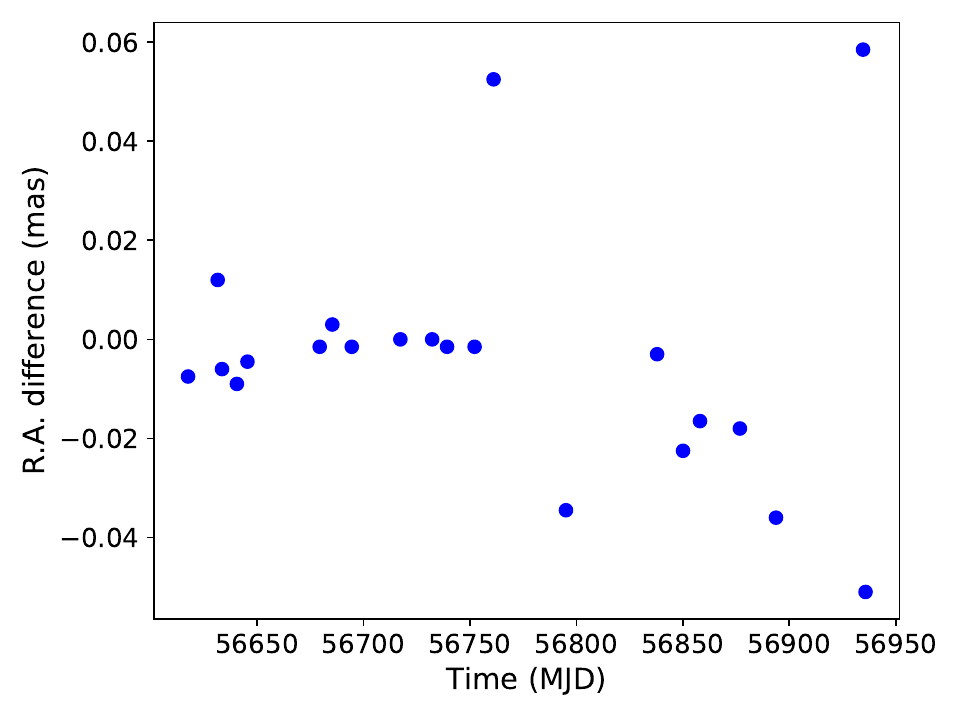}
    \includegraphics[width=0.49\linewidth]{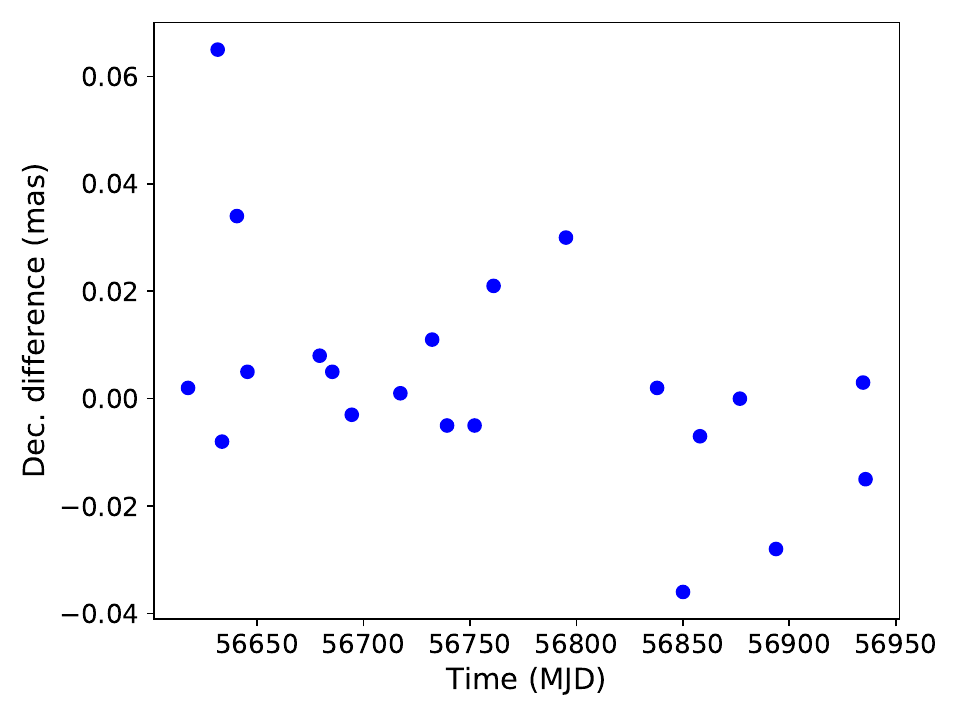}
    \includegraphics[width=0.49\linewidth]{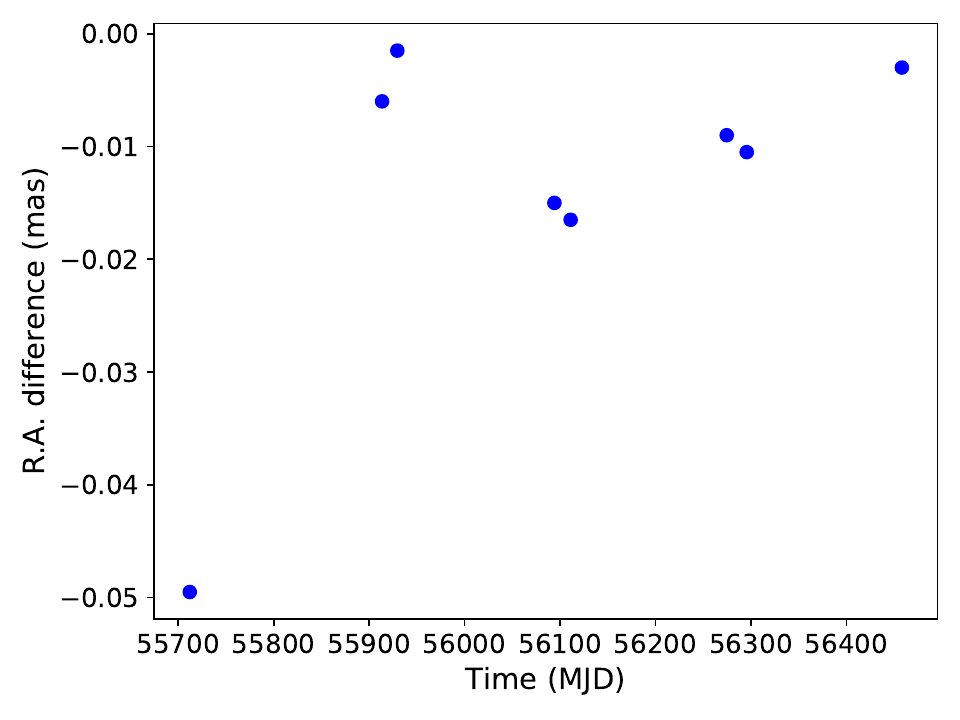}
    \includegraphics[width=0.49\linewidth]{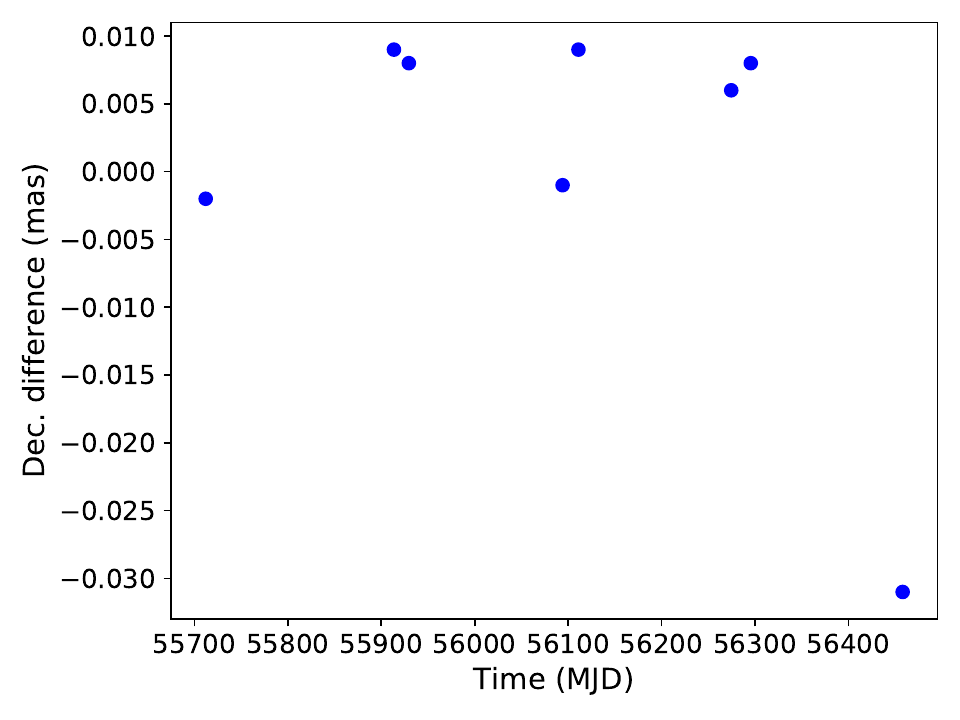}
    \caption{The plots show the B1133+16 position-time series difference (P$_{\rm trad}$ - P$_{\rm Petrov23}$), when the data have been processed using the traditional and Petrov23 mapping functions. For details, see the caption of Figure \ref{fig:pos_diff_J0332}.}
    \label{fig:pos_diff_J1136}
\end{figure*}

\section{Impact of frequency variations on astrometric parameters}\label{sec:6}
To probe the ionospheric propagation effects on astrometry, we have analysed 1.4\,GHz and 1.6\,GHz data sets independently. For each source, the fitted reference epoch positions have an offset at the 1.4\,GHz and 1.6\,GHz (Table \ref{table:ast_results_56300}), which remain almost the same when switching between different TEC mapping functions. It seems this position offset might be due to the mis-modelled structure of the IBC and the residual ionospheric error, which are associated with the ionospheric fluctuations within the spatial and temporal resolution of the global TEC maps.

For each pulsar, the ratio of the uncertainties to the parallax at both bands is close to the ratio of the wavelength squared (almost 1.5; Table \ref{table:ast_results_56300}), which one can expect if the ionosphere is the dominant contributor to the position uncertainties. This underscores the notion that the dominant systematic error in position arises from ionospheric effects at low frequencies. A comparison of astrometric parameters at 1.4\,GHz and 1.6\,GHz is presented in Table \ref{table:ast_results_56300}, confirming the anticipated superiority of astrometric results at 1.6\,GHz, irrespective of the TEC mapping function. 

\section{Conclusion}\label{sec:7}
Our refined astrometric estimates for B0329+54 and B1133+16, based on combined data sets (BD174 and BD152), are shown in Table \ref{table:ast_results_56300}. Including BD174 observations along with the BD152 observations of our pulsars has improved the precision of the proper motion by more than a factor of two, and the parallax precision is enhanced by $\sim73\%$ for each pulsar, compared to \cite{Deller_2019} estimates, which were based solely on BD152 observations. The new astrometric estimates for these pulsars are consistent with \cite{Deller_2019} estimates. The number of BD174 observations is ambitious, but choosing a few observations at regular intervals (a few observations from the beginning, middle, and end) should yield comparable results.

The best-fitted astrometric parameters from all three fitting techniques are consistent within their error bars (Table \ref{table:ast_results_56000}). The least-squares fitting is straightforward but highly sensitive to uncertainties in the input data. In contrast, the bootstrap and Bayesian methods offer more robust parameter and associated uncertainty estimates, particularly when uncertainties in the input data are not well known, which is often the case in astrometry. The Bayesian approach, in particular, provides more accurate parameter estimates if reliable prior information on the fitting parameters is available.

The total error budget for the astrometric measurements is well appreciated, but quantifying the contributions from several factors is challenging. These include: (1) thermal noise, which can be directly obtained from the image and is inversely proportional to the resolution and S/N on the target; (2) differential atmospheric propagation errors, which depend on the angular separation between the target and phase calibrator, the time gap between their scans (in the case of the phase referencing), and the prevailing ionospheric conditions—worsening during periods of increased solar activity-- all these are well known but difficult to measure their effects directly; and (3) structural evolution and core-shift of the primary calibrator. The impact of the second factor can be mitigated by using a strong calibrator located near the target when available. Alternatively, a two-dimensional interpolation technique \citep{Rioja_2017} can be applied if three calibrators enclose the target, though finding such a calibrator combination is often difficult. Recently, \cite{ding2024pinpt} introduced a new technique, so-called \texttt{PINPT}, which combines two-dimensional interpolation and multi-frequency observations to estimate core-shift, achieving millisecond pulsar (J2222-0137) position with 0.2\,mas precision using VLBI. However, this approach presents practical challenges, including a scarcity of reliable calibrators and the increased telescope time and resources required for multi-frequency observations and their analysis. 

The BD174 data set has broadband observations of our pulsars, which provides a unique opportunity to scrutinise systematic uncertainty contributions to astrometric parameters. In Section \ref{sec:6}, we have demonstrated that the primary source of systematic error in L-band pulsar astrometry originates from the ionosphere, with a mitigated impact observed at higher frequency observations. However, observing pulsars at higher frequencies is only feasible with highly sensitive arrays due to their steep spectra. Further, we have examined the impact of the Petrov23 mapping function on astrometry. Our comparative astrometric analysis in Section \ref{sec:5} shows that in the case of in-beam calibration, the small-scale spatial and temporal variations in the ionosphere seem to be more significant than the difference in the TEC mapping functions, which can be more significant at low elevation observations. The difference in the measured pulsar position-time series obtained using traditional and Petrov23 mapping functions (Figures \ref{fig:pos_diff_J0332} and \ref{fig:pos_diff_J1136}) is within the expectation from the position difference estimated using each TEC mapping function (Figure \ref{fig:map_fns}). Additionally, the astrometric parameters remain consistent whether the data is calibrated using the traditional or Petrov23 mapping function (Table \ref{table:ast_results_56300}). However, the uncertainties in fitted parameters slightly increased when using the Petrov23 mapping function (Table \ref{table:ast_results_56300}). The reduced chi-square values from the least squares fitting for different dataset combinations remain almost unchanged, making it unclear that the new mapping function is an improvement for astrometry. Nevertheless, the Petrov23 mapping function more accurately represents realistic situations, as it models the ionosphere as a thick layer, leading to better compensation for dispersive delays. However, the difference is not pronounced in our analysis. At lower elevations—where radio observations are less common— the traditional mapping function would introduce significant errors, and the Petrov23 mapping function might provide better ionospheric delay compensation.

\begin{acknowledgements}
We thank the anonymous referee for the review and helpful comments, which significantly improved the manuscript. Ashish thanks Hao Ding for providing the Python scripts, which are used to estimate the systematic error in pulsar positions. A.K. also expresses gratitude to Avinash Deshpande for the insightful discussions, and appreciates the NRAO (National Radio Astronomy Observatory) help desk for their prompt assistance with queries about certain AIPS tasks and for addressing concerns regarding high noise level in the S-band. The authors thank Gemma Janssen and Ben Stappers for their assistance with predicting the pulse time of arrival for pulsar gating at the VLBA correlator. This study is based on astrometric observations using the VLBA, which is operated by the NRAO. The NRAO is a facility of the National Science Foundation (NSF) operated under a cooperative agreement by Associated Universities, Inc. J.M. acknowledges financial support from the grant CEX2021-001131-S funded by MCIN/AEI/ 10.13039/501100011033 and grant PID2023-147883NB-C21 funded by MCIN/AEI/ 10.13039/501100011033 and by ERDF/EU.

This study utilised several Python packages, including numpy \citep{Harris20}, scipy \citep{2020SciPy-NMeth}, astropy \citep{Astropy-Collaboration13, Astropy-Collaboration18, Astropy-Collaboration22}, matplotlib \citep{Hunter:2007}, bilby \citep{bilby_paper}, psrqpy \citep{psrqpy}, and corner \citep{corner}.

\end{acknowledgements}
\printendnotes
\printbibliography
\end{document}